\documentclass[usenatbib]{mnras}
\usepackage{pdflscape}
\usepackage{color}
\usepackage{IEEEtrantools} %
\usepackage{amsmath} %
\usepackage{amssymb} %
\usepackage{bm} %
\usepackage{upgreek} %
\usepackage[pdftex]{graphicx} %
\usepackage{multirow}
\usepackage{textcomp}
\usepackage{CJK}
\usepackage{rotating}
\usepackage{txfonts}
\usepackage{footnote}
\usepackage[dvipsnames]{xcolor}

\def\aj{AJ}

\def\araa{ARA\&A}
\def\apj{ApJ}

\def\apjs{ApJS}

\def\aap{A\&A}
\def\aapr{A\&A Rev.}
\def\aaps{A\&AS}

\def\mnras{MNRAS}

\def\pra{Phys.~Rev.~A}

\def\ssr{Space Sci.~Rev.}

\def\nat{Nature}

\newcommand{\sect}[1]{\text{Sect.~\ref{#1}}}
\newcommand{\fig}[1]{\text{Fig.~\ref{#1}}}
\newcommand{\tab}[1]{\text{Table~\ref{#1}}}

\newcommand{\marcs}{\textsc{marcs}}
\newcommand{\sme}{\textsc{sme}}
\newcommand{\mist}{\textsc{mist}}

\newcommand{\kms}{\mathrm{km\,s^{-1}}}
\newcommand{\teff}{T_{\mathrm{eff}}}
\newcommand{\lgg}{\log{g}}

\newcommand{\vsini}{\varv\sin\iota}
\newcommand{\vmic}{\xi}
\newcommand{\vmac}{\varv_{\mathrm{mac}}}
\newcommand{\feh}{\mathrm{\left[Fe/H\right]}}
\newcommand{\xh}[1]{\mathrm{\left[#1/H\right]}}
\newcommand{\xfe}[1]{\mathrm{\left[#1/Fe\right]}}

\newcommand{\dex}{\mathrm{dex}}
\newcommand{\Gyr}{\mathrm{Gyr}}
\newcommand{\halpha}{\mathrm{H\upalpha}}
\newcommand{\hbeta}{\mathrm{H\upbeta}}
\title[Verifying abundance
trends in the open cluster M67 using non-LTE modelling]{The GALAH Survey: 
Verifying abundance trends in the open cluster M67
using non-LTE modelling} 
\author[X.~D.~Gao et al.]{Xudong~Gao$^{1,2}$\thanks{E-mail:gao@mpia.de}, 
Karin~Lind$^{1,3}$, Anish~M.~Amarsi$^{1}$, Sven~Buder$^{1,2}$, Aaron~Dotter$^{4}$, \newauthor
Thomas~Nordlander$^{5,6}$, Martin~Asplund$^{5,6}$, Joss~Bland-Hawthorn$^{7,6,8}$, \newauthor
Gayandhi~M.~De~Silva$^{9,6}$, Valentina~{D'Orazi}$^{10}$, Ken~C.~Freeman$^{5}$, Janez~Kos$^{7}$, \newauthor
Geraint~F.~Lewis$^{7}$, Jane~Lin$^{5,6}$, Sarah~L.~Martell$^{11,6}$, Katharine.~J.~Schlesinger$^{5}$,\newauthor
Sanjib~Sharma$^{7,6}$, Jeffrey~D.~Simpson$^{11}$, Daniel~B.~Zucker$^{12,9}$, Toma\v{z}~Zwitter$^{13}$, \newauthor
Gary~Da~Costa$^{5}$, Borja~Anguiano$^{14}$, Jonathan~Horner$^{15}$, Elaina~A.~Hyde$^{16}$, \newauthor
Prajwal~R.~Kafle$^{17}$, David~M.~Nataf$^{18}$, Warren~Reid$^{16,12}$, Dennis~Stello$^{12,19,6}$,\newauthor
Yuan-Sen~Ting$^{20,21,22}$, and~the~GALAH~collaboration\\
$^{1}$Max Planck Institute f\"ur Astronomy, K\"onigstuhl 17, D-69117 Heidelberg, Germany\\
$^{2}$Fellow of the International Max Planck Research School for Astronomy \& Cosmic Physics at the University of Heidelberg, Germany\\
$^{3}$Department of Physics and Astronomy, Uppsala University, Box 516, SE-751 20 Uppsala, Sweden\\
$^{4}$Harvard-Smithsonian Center for Astrophysics, Cambridge, MA 02138, USA\\
$^{5}$Research School of Astronomy $\&$ Astrophysics, Mount Stromlo Observatory, Australian National University, ACT 2611, Australia\\
$^{6}$Center of Excellence for Astrophysics in Three Dimensions (ASTRO-3D), Australia\\
$^{7}$Sydney Institute for Astronomy (SIfA), School of Physics, A28, The University of Sydney, NSW, 2006, Australia\\
$^{8}$Miller Professor, Miller Institute, University of California Berkeley, CA 94720, USA\\
$^{9}$Australian Astronomical Observatory, 105 Delhi Rd, North Ryde, NSW 2113, Australia\\
$^{10}$Istituto Nazionale di Astrofisica, Osservatorio Astronomico di Padova, vicolo dell'Osservatorio 5, 35122, Padova, Italy \\
$^{11}$School of Physics, University of New South Wales, Sydney, NSW 2052, Australia\\
$^{12}$Department of Physics and Astronomy, Macquarie University, Sydney, NSW 2109, Australia\\
$^{13}$Faculty of Mathematics and Physics, University of Ljubljana, Jadranska 19, 1000 Ljubljana, Slovenia\\
$^{14}$Department of Astronomy, University of Virginia, P.O. Box 400325 Charlottesville, VA 22904-4325, USA\\
$^{15}$University of Southern Queensland, Toowoomba, Queensland 4350, Australia\\
$^{16}$Western Sydney University, Locked Bag 1797, Penrith South DC, NSW 1797, Australia\\
$^{17}$ICRAR, The University of Western Australia, 35 Stirling Highway, Crawley, WA 6009, Australia\\
$^{18}$Department of Physics and Astronomy, The Johns Hopkins University, Baltimore, MD 21218, USA\\
$^{19}$Stellar Astrophysics Centre, Department of Physics and Astronomy, Aarhus University, DK-8000, Aarhus C, Denmark\\
$^{20}$Institute for Advanced Study, Princeton, NJ 08540, USA\\
$^{21}$Department of Astrophysical Sciences, Princeton University, Princeton, NJ 08544, USA\\
$^{22}$Observatories of the Carnegie Institution of Washington, 813 Santa Barbara Street, Pasadena, CA 91101, USA\\}

\begin{document}
\begin{CJK*}{UTF8}{gbsn}
\date{Accepted ---. Received ---; in original form ---}
\pagerange{\pageref{firstpage}--\pageref{lastpage}} \pubyear{---}

\maketitle 
\end{CJK*}
\label{firstpage}
\begin{abstract}

Open cluster members are coeval and share the same initial bulk chemical
composition. Consequently,
differences in surface abundances between members of a cluster
that are at different evolutionary stages 
can be used to study the effects of 
mixing and internal chemical processing.
We carry out an abundance analysis of seven elements (Li, O, Na, Mg, Al, Si, Fe) in 66 stars
belonging to the open cluster M67, based on high resolution GALAH spectra,
1D \marcs~model atmospheres, and non-local
thermodynamic equilibrium (non-LTE) radiative transfer. 
From the non-LTE analysis,
we find a typical star-to-star scatter in the abundance ratios of 
around $0.05\,\dex$.
We find trends in the abundance ratios with effective temperature,
indicating systematic differences in the surface abundances
between turn-off and giant stars;
these trends are more pronounced when LTE is assumed.
However, trends with effective temperature remain significant for 
Al and Si also in non-LTE.
Finally, we compare the derived abundances with prediction from stellar evolution models
including effects of atomic diffusion. We find overall good agreement for the abundance patterns of 
dwarfs and subgiant stars, but the abundances of cool giants 
are lower relative to less evolved stars than predicted by the diffusion models,
in particular for Mg.

\end{abstract}

\section{Introduction}
\label{sect:introduction}

Under the assumption that open clusters formed in a single burst of
star formation from a chemically homogeneous and well-mixed progenitor cloud
(e.g. \citealt{2006AJ....131..455D, 2007AJ....133.1161D, 2010A&amp;A...511A..56P, 2014A&amp;A...563A..44M, 2014Natur.513..523F})
open cluster members are coeval, and share the same initial bulk chemical
compositions, differing only in their initial stellar masses.  Based
on the chemical homogeneity in star clusters, the chemical tagging technique, as proposed by
\citet{2002ARA&amp;A..40..487F}, can be used to reconstruct stellar groups
that have dispersed.  For example, \citet{2018MNRAS.473.4612K} have
successfully identified two new members of the Pleiades, located far from
the cluster centre, with chemical tagging, and recovered seven observed
clusters in chemical space by using t-distributed stochastic neighbour
embedding (t-SNE). To study Galactic Archaeology by chemical tagging, a
large amount of high quality observed data will be provided by massive high
resolution spectroscopic surveys such as the GALactic Archaeology with HERMES
(GALAH) \citep{2015MNRAS.449.2604D}, Gaia-ESO \citep{2012Msngr.147...25G} and
APOGEE \citep{2017AJ....154...94M}, WEAVE \citep{2012SPIE.8446E..0PD}, 
4MOST \citep{2012SPIE.8446E..0TD}.

However, recent studies have demonstrated that, in the same open
cluster, the surface abundances measured in (unevolved) dwarf stars are
apparently offset compared to those measured in (evolved) giant stars
\citep[e.g.][]{2009A&amp;A...504..845V,
2009ApJ...701..837S,2014A&amp;A...562A.102O, 2017MNRAS.466..613M}. These
trends with evolutionary stage cannot be explained by the simple
stellar evolution model, in which convection is the only internal
mixing process. 

Atomic diffusion is one possible explanation for these surface
abundance trends \citep{1984ApJ...282..206M}.  Atomic diffusion can
perturb the surface abundances of dwarfs with shallow
convection zones: different chemical species will be underabundant or
overabundant to varying degrees in the stellar atmosphere, depending on the
competing effects of gravitational settling and radiative acceleration.
Furthermore, once the star leaves the turn-off point and starts climbing
the red giant branch, the deeper convection zone will restore the original
composition in the atmosphere.

Systematic abundance trends with evolutionary stage have also been
measured in a number of globular clusters,
which can be well described by using atomic diffusion models 
with extra turbulent mixing below the convection zone
\citep[e.g.][]{2007ApJ...671..402K, 2009A&amp;A...503..545L,
2012ApJ...753...48N, 2014A&amp;A...567A..72G, 2016A&amp;A...589A..61G}.
However, these globular clusters are old, and only probe the low metallicity
regime ($-2.3<\mathrm{[Fe/H]}<-1.5$). They also show anti-correlations
in some light elements, which is thought to be produced by intra-cluster
pollution by short-lived stars of the first cluster generation
\citep[e.g.][]{2006A&amp;A...458..135P}.  In contrast, open clusters
probe the metallicity and age range typical of the Galactic disk, and
are not expected to have experienced such internal pollution. Thus, the stellar
surface compositions of open cluster members should truly reflect the
primordial abundances from the proto-cluster, unless they have been altered
by evolutionary effects.

M67 is an ideal target to study such phenomena with a well determined reddening
($\mathrm{E(B-V)=0.041}$) and distance modulus 
($\mu = 9.70-9.80$; \citealt{2009ApJ...698.1872S,2009A&amp;A...503..165Y}),
which permits a detailed
spectroscopic study of even its main sequence stars.  M67 has been widely
studied, with the various studies obtaining slightly different results. For example,
the averaged metallicities ([Fe/H]) ranges from
$-0.04$~to $+0.05$~\citep{1991AJ....102.1070H,2000A&amp;A...360..499T,
2005AJ....130..597Y,2006A&amp;A...450..557R,
2008A&amp;A...489..677P,2008A&amp;A...489..403P}, while determinations
of the cluster age vary between $3.5$~to $4.8\,\Gyr$~\citep{2008A&amp;A...484..609Y,
2011A&amp;A...528A..85O}.  Considering the uncertainties in the different
studies, they are all consistent with the conclusion that chemical composition 
and age of M67 are similar to those of the Sun.
It has even been suggested that this is the original birthplace of the Sun
\citep{2011A&amp;A...528A..85O}, but this has been challenged
\citep{2012AJ....143...73P,2016A&amp;A...593A..85G}.  

Previous studies of abundance trends in M67 have been based on small
samples \citep[e.g.][]{2000A&amp;A...360..499T,
2005AJ....130..597Y,2006A&amp;A...450..557R,2008A&amp;A...489..403P,
2010A&amp;A...511A..56P}.  In particular, \citet{2014A&amp;A...562A.102O}
found that heavy element abundances in dwarf stars are reduced by
typically $0.05\,\dex$~or less, compared to those in sub-giants.  
Atomic diffusion has already been suggested as the underlying cause for the
abundance trends in M67
\citep{2014A&amp;A...562A.102O,2017MNRAS.466.2161B,2018ApJ...857...14S}; 
we note that, for the mass range of M67 (less than about $2\,\rm{M}_{\odot}$),
intermediate and heavy elements will not be influenced by 
nuclear reactions associated with dredge-up
\citep{2016A&amp;A...589A.115S}; the light elements Li, Be, and B can
be destroyed during the course of the first dredge-up.

However, in order to use the surface abundance trends to make
quantitative statements about atomic diffusion processes, the
measured surface abundances must be accurate. To date, most abundance
analyses have employed the simplifying assumption of local
thermodynamic equilibrium (LTE) for the gas in the stellar atmosphere.
In reality, conditions in the line-forming regions are such that
radiative transitions typically dominate over collisional
transitions; the non-thermal radiation field thus drives the gas away from
LTE. Thus, to measure surface abundances to better than $0.05\,\dex$,
departures from LTE must be taken into account \citep[e.g.][and references
therein]{2005ARA&amp;A..43..481A}.  Moreover, the errors arising from the
assumption of LTE are systematic, and can therefore result in spurious
abundance trends which, if taken to be real, can lead to incorrect
conclusions about stellar and Galactic physics.
For example, recent studies in open clusters show a remarkable enhancement 
of Na abundance compared with field stars, however, this Na-enhancement 
is only an artefact of non-LTE effects, which have been shown by 
\cite{2015MNRAS.446.3556M}.

Here we present a detailed non-LTE abundance analysis of lithium, oxygen,
sodium, magnesium, aluminium, silicon, and iron, across $66$~M67 members.
We employ a homogeneous data set drawn from GALAH survey 
\citep{2015MNRAS.449.2604D}, to study how departures
from LTE can influence the observed abundance trends in M67.  By comparing
the trends against recent stellar models that include atomic diffusion, we
investigate how departures from LTE influence interpretations about the
efficiency of mixing processes in stellar atmospheres.

The rest of paper is structured as follows.  In \sect{sect:data}~we describe
the observational data used in this study and the sample selection.  In
\sect{sect:analysis}~we describe the abundance analysis.  In
\sect{sect:results}~we present the inferred abundances and consider the
abundance trends and the non-LTE effects.  In \sect{sect:discussion}~we
discuss these results in relation to others in the literature, as well as to
different models of stellar mixing.  Our conclusion are presented in \sect{sect:conclusion}.

\section{Observational data and sample selection}
\label{sect:data} 

\begin{figure} 
\includegraphics[scale=1.0,width=\columnwidth]{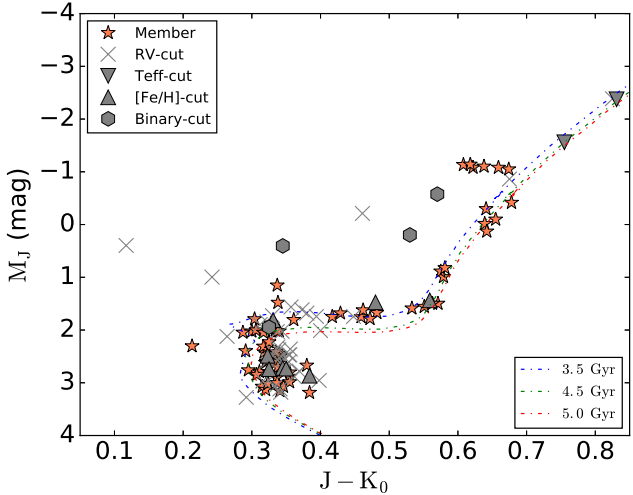}
\caption{Colour-Magnitude Diagram of the open cluster M67
generated by using the
photometric data from 2MASS \citep{2006AJ....131.1163S} with a
distance modulus of 9.70 and reddening $\mathrm{E(B-V)=0.041}$ mag. The excluded
stars are represented by different grey symbols for different selection
processes.  The cluster members selected and used in this study are marked
as filled red star symbols. The spectroscopic binaries found in our final
sample are shown as grey hexagon. Solar abundance isochrones corresponding to an
age of $3.5\,\Gyr$, $4.5\,\Gyr$ and $5.0\,\Gyr$ are shown as
dot-dashed lines in different colours.} 
\label{fig:CMD}
\end{figure}

\begin{figure}
\includegraphics[scale=1.0,width=\columnwidth]{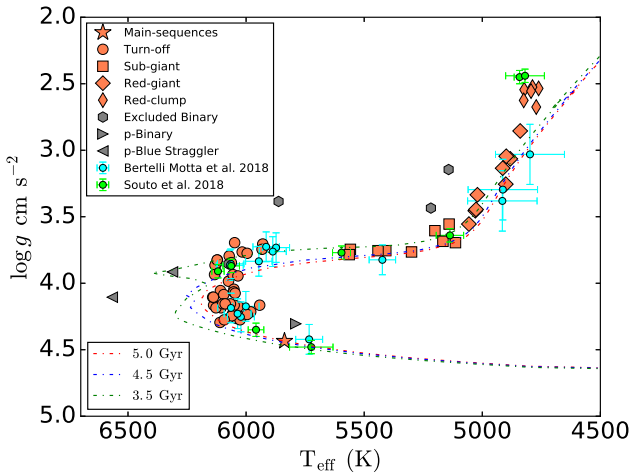}
\caption{Theoretical isochrones of M67 with solar metallicity and
different ages. The sample stars are divided into main-sequence, turn-offs, sub-giants and
giants represented by different symbols. The excluded binaries, possible
blue stragglers and unresolved binary are 
also displayed. The effective temperature and gravity of the
targeted stars has been offset by $59\,\mathrm{K}$ and
$0.22\,\dex$, respectively. Results from \citet{2018MNRAS.478..425B} and \citet{2018ApJ...857...14S}
are also shown for comparsion.}
\label{fig:isochrone} 
\end{figure}

\begin{figure*}
\includegraphics[scale=1.0,width=\columnwidth]{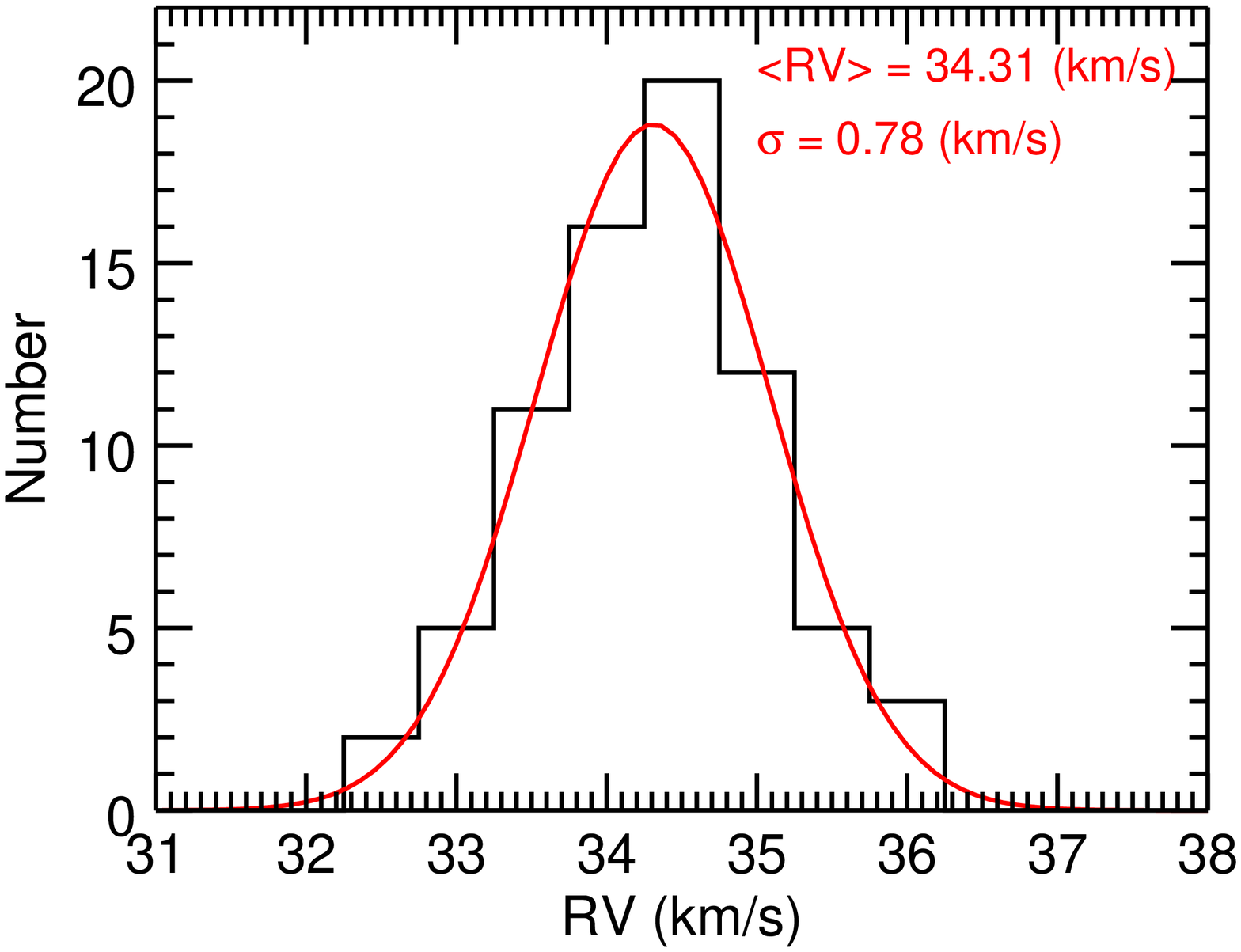}
\includegraphics[scale=1.0,width=\columnwidth]{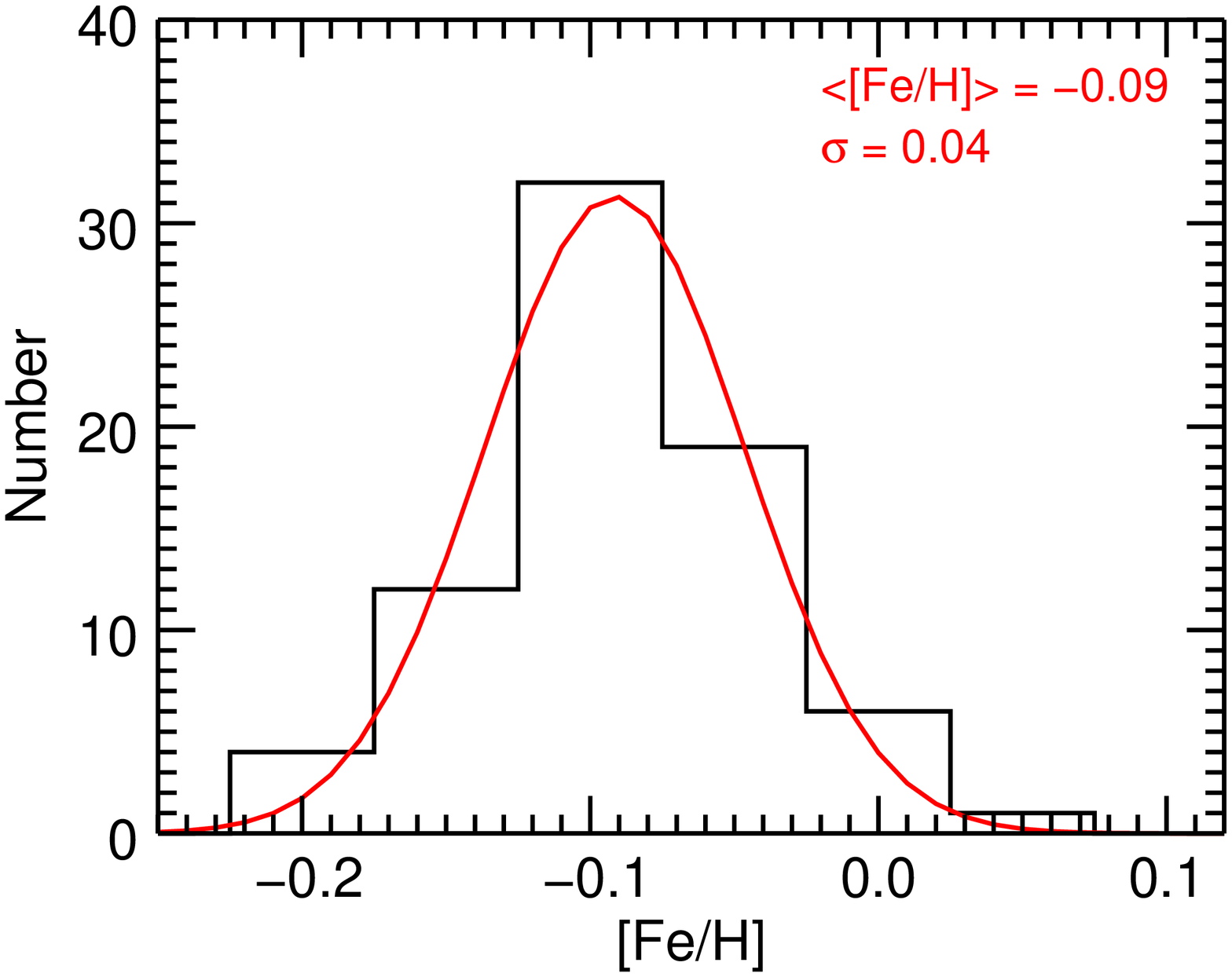}
\caption{Histogram of the radial velocity and metallicity distributions
of the final members selected in M67. The corresponding Gaussian fit to
the distributions are also been shown in red lines.}
\label{fig:radv_metallicity}
\end{figure*}

The spectroscopic observations of target stars in M67 were taken from the
GALAH survey, whose main science goal is to reveal the formation and
evolutionary history of the Milky Way using chemical tagging
\citep{2002ARA&amp;A..40..487F}.  The stars in the GALAH survey were
observed with the HERMES spectrograph
\citep{2015JATIS...1c5002S} mounted on the Anglo-Australian Telescope (AAT).
The spectra provided by HERMES are in fixed format with four noncontiguous
wavelength bands, $471.3$-$490.3\,\mathrm{nm}$~(Blue),
$563.8$-$587.3\,\mathrm{nm}$~(Green), $647.8$-$673.7\,\mathrm{nm}$~(Red),
and $758.5$-$788.7\,\mathrm{nm}$~(IR).

HERMES is designed to operate at two resolution modes for GALAH.
During the normal operation, HERMES observes with a resolving powers of $R\sim28,000$.
A higher resolution of $R\sim42,000$ was used during part of the GALAH pilot survey 
\citep{2017MNRAS.465.3203M}.  
This study is based only on spectra taken
in the higher resolution mode (i.e.~$R\sim42,000$).
The observations were carried out during the period of 7-14~February 2014.
The exposure time ranges from $3600\,\mathrm{s}$ to $7200\,\mathrm{s}$.
The spectra were reduced using the dedicated GALAH reduction pipeline
\citep{2017MNRAS.464.1259K}, 
with 2dfdr and IRAF used to perform bias subtraction, flat
fielding, wavelength calibration and spectral extraction.
The sky background was subtracted from each individual observation.
Observed spectra of the same object with different observation dates were
stacked for higher signal-to-noise (SNR). All the targets satisfy
$\mathrm{SNR}>50$~in Green, Red and IR arms.
 
In \fig{fig:CMD} we show the colour-magnitude diagram (CMD) for the observed 
M67 sample (stars with $8.8<V<14$). The original candidate list was sourced from
the precise optical photometry available from Stetson's database of photometric 
standard fields
\footnote{\url{http://www.cadc-ccda.hia-iha.nrc-cnrc.gc.ca/en/community/STETSON/standards/}}.
\fig{fig:CMD} shows the $\mathrm{M_{J}}$, $\mathrm{(J-K)_{0}}$ CMD for the stars using 
the Two Micron All Sky Survey photometry \citep{2006AJ....131.1163S} 
with a M67 distance modulus of 9.70 and reddening $\mathrm{E(B-V)=0.041}$ mag.
We determined the radial velocities and spectroscopic stellar parameters as
described in \sect{sect:analysis_parameters}.  To refine the membership
selection, we iteratively rejected $2\,\sigma$~outliers in radial velocity.
We also excluded two probable members that are cooler than
$4500\,\mathrm{K}$, since our approach to determining spectroscopic
parameters (based on unblended H and Fe lines) is not reliable 
at these cool temperatures.  Finally, we
retained all the stars within $3\,\sigma$ in [Fe/H] as our final sample,
thereby rejecting another 8 probable foreground objects of similar radial
velocity as the cluster.  The abandoned and retained stars are shown in
different symbols in \fig{fig:CMD}.
	
In \fig{fig:radv_metallicity} we show histograms of the radial velocity
and metallicity distributions of the final sample of stars, together with a
Gaussian fit with $\mathrm{<RV> = 34.31\,\kms}$ and $\sigma =0.78\,\kms$, which is
consistent with the result from \citet{2015AJ....150...97G}
($\mathrm{RV = 33.64 \pm 0.96\,\kms}$).  We also made a cross-match of our
targeted stars in SIMBAD \citep{2000A&amp;AS..143....9W} by 
using the coordinates to identify the
corresponding objects within a radius of 2 arcsec.  We found four stars in
our final sample (marked as grey hexagon in \fig{fig:CMD}) that are listed 
as spectroscopic binaries in SIMBAD;
we excluded these binaries in the sample.
Furthermore, by checking the positions of all the left stars in the isochrones (see \fig{fig:isochrone}),
we excluded two stars that could be blue stragglers whose temperature are
significantly hotter than the other turn-off stars. The coolest dwarf that might well be an unresolved 
binary has been removed, which lies well above the isochrones in $\log{g}$.
We also see that six stars
stand out in \fig{fig:CMD} as likely red clump stars.  
The final stellar sample contains 
$66$~stars with high resolution spectra, including
turn-off, sub-giant, red-giant, and red-clump stars, 
as well as a single solar-like main-sequence star.

\section{Abundance analysis} 
\label{sect:analysis}


The spectra were analysed using a modified version of the GALAH analysis
pipeline, which is developed for a full scientific analysis of the GALAH
survey and has been applied to determine stellar parameters and abundances
in a number of recent studies 
\citep[e.g.][]{Sharma2017, Wittenmyer2017,
2018arXiv180101514D}. The pipeline and the
results for the full survey sample are further described and
evaluated in GALAH's second data release
paper \citep{2018MNRAS.478.4513B}. 
The input data for this pipeline includes: the
reduced observed spectra and corresponding measurement errors
(\sect{sect:data}); initial guesses for the stellar atmosphere parameters
and radial velocity; 
reference solar abundances; and a list of atomic and
molecular lines. 
The spectra, which have been radial velocity corrected as
described in \citet{2017MNRAS.464.1259K}, were first continuum-normalised
using straight lines over $3$-$60\,\mathrm{\AA}$ wide segments in all four
arms.  Wavelength regions contaminated by telluric or sky lines were removed
\citep{2018MNRAS.478.4513B}.  The radiative transfer and abundance analysis was
carried out using the automated spectrum analysis code \textsc{Spectroscopy
Made Easy} \citep[{\sme};][]{2017A&amp;A...597A..16P} We detail aspects of
this pipeline in the remainder of this section.

\subsection{Atmosphere grids}
\label{sect:analysis_atmosphere}

The spectral line synthesis with \sme~is based on \marcs~model atmospheres
\citep{2008A&amp;A...486..951G} with atmospheric parameters spanning
effective temperatures $2500\leq\teff/\mathrm{K}\leq8000$, surface gravities
$-0.5\leq\log_{10}\left(g / \mathrm{cm\,s^{-2}}\right)\leq5.0$, and
metallicities $-5.0\leq\feh\leq1.0$.  Spherical models were used for
$\lgg\leq3.5$ and plane-parallel models were otherwise used.  The standard
chemical composition grid was adopted, which uses the solar chemical
composition of \citet{2007SSRv..130..105G}, scaled by $\feh$, and with an
enhancement to $\upalpha$-elements of $0.1\,\dex$ for $\feh=-0.25$,
$0.2\,\dex$ for $\feh=-0.5$, $0.3\,\dex$ for $\feh=-0.75$, and $0.4\,\dex$
for $\feh\leq-1.0$.

\subsection{Non-LTE grids}
\label{sect:analysis_nlte}

For non-LTE calculations in \sme, instead of solving the non-LTE radiative
transfer equations directly, grids of pre-computed departure coefficients
$\beta = n_\mathrm{NLTE}/n_\mathrm{LTE}$ as functions of optical depth 
were employed instead, as described in
\citet{2017A&amp;A...597A..16P}. When performing the 
spectral fitting for stellar parameter determinations,
as well as the spectral fitting for chemical abundance determinations,
the grids of pre-computed departure coefficients (for
each stellar model and target abundance) were read in and interpolated based
on a given stellar model and non-LTE abundance.
Then the corresponding departure coefficients were
applied to the corresponding LTE level populations to synthesise the lines.

The non-LTE departure coefficient grids for all the elements were taken from
recent non-LTE radiative transfer calculations based on 1D hydrostatic model
\marcs~atmospheres (i.e.~consistent with the rest of the analysis).
The calculations themselves, and/or the model atoms,
were presented in the following studies:
\begin{itemize} 
\item lithium: \citet{2009A&amp;A...503..541L} 
\item oxygen: \citet{2015MNRAS.454L..11A} (model atom)
\item sodium: \citet{2011A&amp;A...528A.103L} 
\item magnesium: \citet{2016A&amp;A...586A.120O} 
\item aluminium: \citet{2017arXiv170801949N} 
\item silicon: \citet{2017MNRAS.464..264A} (model atom)
\item iron: \citet{2016MNRAS.463.1518A}
\end{itemize} We refer the reader to those papers for details on the model
atoms; we only provide a brief overview here.

Energy levels and radiative data were taken from various databases, as
appropriate or applicable: NIST \citep{2012AAS...21944301R}, TOPbase
\citep{1988JPhB...21.3669P}, TIPbase \citep{1997A&amp;AS..122..167B}, and
the Kurucz online datebase \citep{1995ASPC...78..205K}.
Inelastic collisional processes, between the
species in question and either free electrons or neutral hydrogen atoms, can
be a major source of uncertainty in non-LTE analyses
\citep[e.g.][]{2016A&amp;ARv..24....9B}.  The oxygen, sodium and magnesium
aluminium grids benefit from X+e inelastic collision data based on the
R-matrix method \citep[e.g.][]{1971JPhB....4..153B,1974CoPhC...8..149B},
while the collision data for aluminium is calculated by using the 
Breit-Pauli distorted wave \citep{2011CoPhC.182.1528B}.
Both methods are more reliable than commonly used alternatives, such as the van
Regemorter recipe \citep{1962ApJ...136..906V}.  

Furthermore, more realistic cross-sections for inelastic collisions with neutral hydrogen
(X+H) are included in the calculations of all the element grids, which is in turn more reliable
than the commonly used Drawin recipe
\citep{1984A&amp;A...130..319S,1993PhST...47..186L}.
To be more specific, for Li, the rate coefficients for inelastic collisions with neutral hydrogen 
were accounted for \citep{2003PhRvA..68f2703B,2003A&amp;A...409L...1B}; 
for O, the rate coefficients were treated by the formula from \cite{1968ZPhy..211..404D} 
with a correction followed by \cite{1993PhST...47..186L};
for Na, the rate coefficients were adopted from \cite{2010A&amp;A...519A..20B}; 
for Mg, the rate coefficients were based on the accurate calculations from \cite{2012A&amp;A...541A..80B};
for Al, the rate coefficients were taken form the computation of \cite{2013A&amp;A...560A..60B}; for Si, the rate coefficients of low and intermediate levels were used from \cite{2014A&amp;A...572A.103B}; for \ion{Fe}{I}, the rate coefficients were calculated with the asymptotic two-electron method, which was applied to Ca+H in 
\cite{2016PhRvA..93d2705B}. 
Since the reactions between \ion{Fe}{II} and hydrogen are not very prominent, thus for \ion{Fe}{II}, 
the collision excitation with hydrogen was still implemented by the old recipe of \cite{1968ZPhy..211..404D}.

\subsection{Spectroscopic stellar parameters}
\label{sect:analysis_parameters}

\begin{figure}
\includegraphics[scale=1.0,width=\columnwidth]{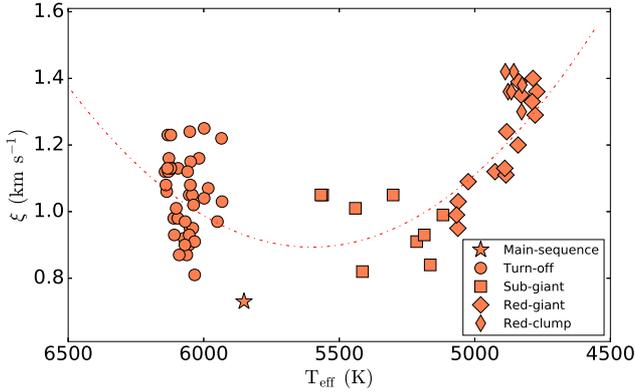}
\caption{Microturbulence $\vmic$ as a
function of effective temperature, when
treated as a free parameter in stellar parameters calculation. 
This distribution was
fitted by an empirical quadratic polynomial,
in order to determine the relation between these two parameters
that was subsequently enforced.}
\label{fig:vmic_teff}
\end{figure}

To avoid degeneracy issues that result from having
too many free model parameters, the analysis separates
the determination of the surface elemental abundances from the rest of the
stellar parameters, namely the atmospheric parameters $\teff$, $\lgg$,
$\feh$, as well as projected rotational velocities $\vsini$, and
line-of-sight radial velocity RV.  In addition,
microturbulence $\vmic$~and macroturbulence $\vmac$~are standard parameters
in 1D atmosphere analysis used to model the impact of convective motions on
the spectral lines \citep[e.g.][Chapter 17]{2005oasp.book.....G}.  
In principle, $\vmic$~could be set as a free parameter when
fitting the spectrum, but we find that this parameter has similar values for similar stars. 
Additionally, macroturbulence and projected rotation $\vsini$ have a degenerate influence 
on spectral line broadening and cannot been disentangled, 
especially for the slowly rotating cool stars. Therefore, in this project we 
applied fixed values for $\vmic$, which are obtained 
from an empirical relation as a function of $\teff$~(see \fig{fig:vmic_teff}),
while we treated $\vsini$ as a free parameter with a rotational broadening profile 
\citep[e.g.][Chapter 18]{2005oasp.book.....G} and set $\vmac$~as zero. 
During this procedure, the synthetic spectra were also
convolved with a Gaussian instrumental profile of varying resolution over
each arm, which is the dominant source of broadening.

The stellar parameters were determined simultaneously, by fitting (via
$\chi^{2}$ minimisation) the observed profiles of \ion{Sc}{I}, \ion{Sc}{II},
\ion{Ti}{I}, \ion{Ti}{II}, \ion{Fe}{I}, and \ion{Fe}{II}~lines that were
unblended and that had reliable atomic data, as well as two of the Balmer
lines: $\halpha$~and $\hbeta$.  
The benefit of this approach is that, for example, both the temperature
sensitive Balmer line wings and the excitation-balance of neutral iron-peak
species strongly influence the effective temperature determination; similar
statements can be made for the inferred surface gravity and metallicity
(\sect{sect:analysis_nlte}).
In this process, iron was treated
in non-LTE \citep{2016MNRAS.463.1518A}, unless otherwise stated. 
The non-LTE effects on iron lines are small,
for late-type stars of solar-metallicity
\citep[e.g.][]{2012MNRAS.427...50L} and
we find this choice has only a small influence on the values of 
the other stellar parameters
(the mean differences in $\teff$~and $\lgg$~under the assumption of LTE and non-LTE 
are $3.5\,\mathrm{K}$ and $0.01\,\dex$, respectively). 

As described in GALAH's second data release paper \citep{2018MNRAS.478.4513B}, 
a bias in surface gravity of 0.15dex and a bias in metallicity of 0.1dex is found for purely spectroscopic SME results when 
compared to results including interferometric, astrometric, and/or asteroseismic constrains. These offsets were applied to 
all survey targets in Buder et al., in a similar fashion to other large spectroscopic surveys, such as 
in APOGEE \citep[][Sect.~5]{2015AJ....150..148H} and 
RAVE \citep[][Sect.~6]{2017AJ....153...75K}.

In this study, we chose to use only the Sun as our reference star, because 
our cluster stars are very close to solar metallicity.  
By analyzing a high resolution solar
spectrum (\sect{sect:analysis_solar}), 
we find that our analysis pipeline requires 
positive offsets in $\teff$, $\lgg$ and $\feh$ of
$59\,\mathrm{K}$, $0.22\,\dex$ and $0.09\,\dex$ respectively, 
to match the reference solar values. We apply these offsets to our spectroscopic parameters
before determining chemical abundances.

Since our sample spans a large range in stellar parameters, 
we could have attempted to design a more sophisticated calibration method involving 
more reference stars. However, our simple method has the advantage of 
preserving the relative parameter differences found by spectroscopy and 
therefore do not strongly influence the derived abundance trends.
Our assumption is thus that the internal precision of our spectroscopic method
is reliable enough to comment on abundance trends.

As a sanity check, in \fig{fig:isochrone} we compare
our effective temperatures
and surface gravities with theoretical cluster isochrones. 
The three stellar evolutionary
tracks and isochrones have been produced using the Padova database
\citep{2012MNRAS.427..127B, 2014MNRAS.444.2525C,2014MNRAS.445.4287T}, with
solar metallicity ($Z=0.0142$), but different ages of $t=3.5\,\Gyr$,
$t=4.5\,\Gyr$~(close to that of the Sun),
and $t=5.0\,\Gyr$.  The parameters of the stars fall
into the reasonable region of the isochrone tracks, without 
any further calibrations.

\subsection{Spectroscopic abundances} 
\label{sect:analysis_abundances}

\begin{figure*} 
\includegraphics[scale=1.0,width=\textwidth]{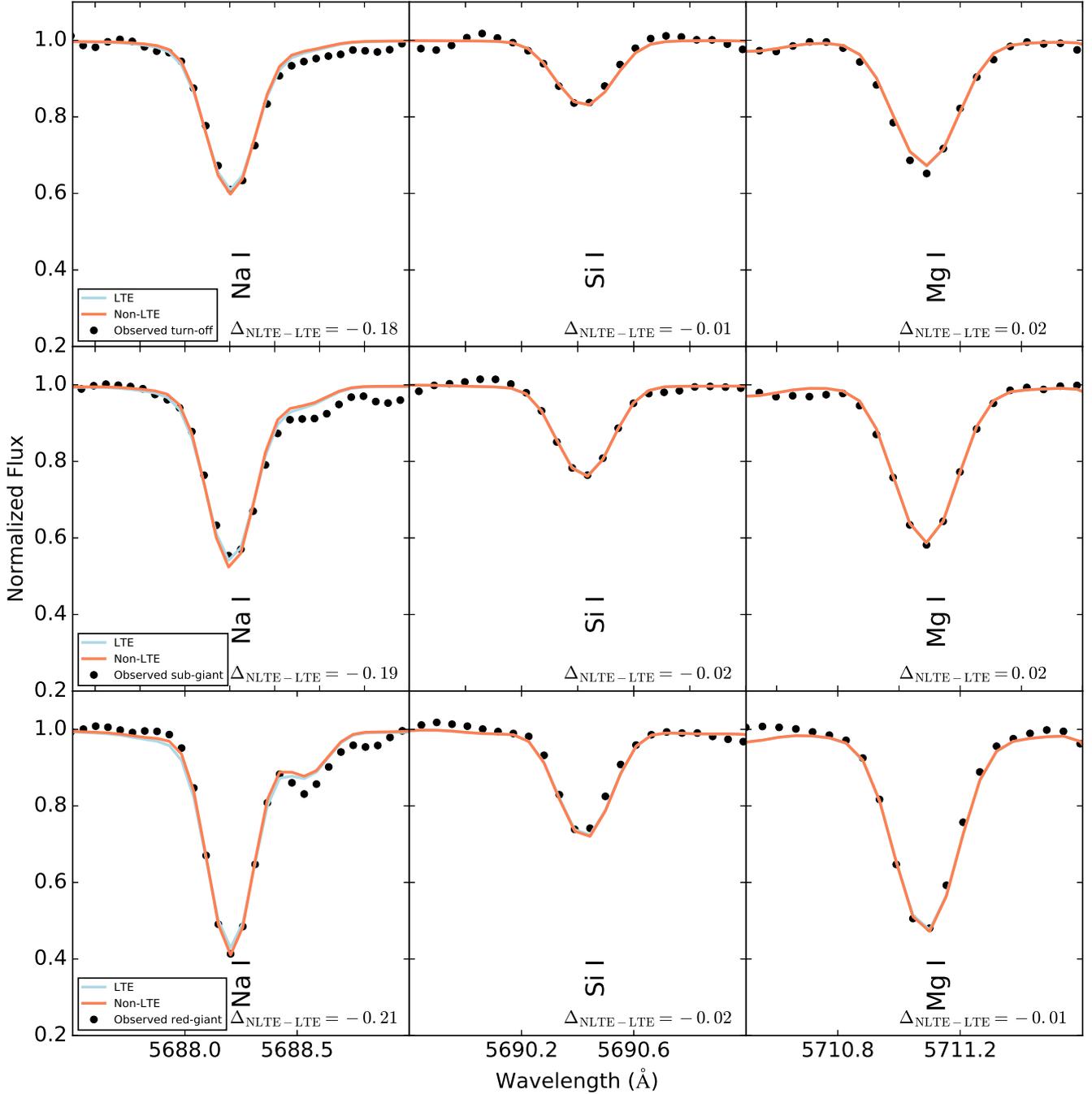} 
\caption{Typical best-fit synthetic LTE and non-LTE line profiles
of Na, Mg and Si compared with the observed spectra 
of three stars in different evolutionary stage; a turn-off, 
a sub-giant and a giant. Only abundance is set as a free parameters 
in these fittings.
Abundance differences between non-LTE 
and LTE synthesis are showed in the labels.}
\label{fig:syn_obs}
\end{figure*}

In principle, GALAH spectra can allow
for up to 30 elements to be determined, but
here we only focus on those for which we have non-LTE grids for.  
Having obtained the
optimal stellar parameters (\sect{sect:analysis_parameters}), elemental
abundances for lithium, oxygen,
sodium, magnesium, aluminium, and silicon were then inferred;
the abundance of iron was also re-inferred, using only
iron lines. The trace element assumption
was employed here: i.e.~that a small change to the abundances of the
particular element being studied has a negligible impact on the background
atmosphere and hence the optimal stellar parameters.  Thus, the stellar
parameters were kept fixed, and the only free parameters were the elemental
abundances. The synthesis of the spectral lines incorporated non-LTE
departure coefficients (\sect{sect:analysis_nlte}).

Unsaturated, unblended lines are preferred as abundance indicators.  For
partially blended lines in the list, synthetic spectra are fitted in an
appropriate selected spectral region that neglects the blended part of the
line.  Likewise, the abundances were calculated from those lines using
$\chi^2$ minimised synthetic fits. All of the lines used in the
abundance analysis and their detailed information are presented in
\cite{2018MNRAS.478.4513B}.
\fig{fig:syn_obs} shows the comparison between observed and best-fit
synthetic line profiles of Na, Mg and Si in both LTE and non-LTE 
for three stars coming from 
different groups: turn-offs, sub-giants and giants.
During these fittings, only abundance is set as a free parameter. 
Abundance difference between non-LTE and LTE synthesis can be substantial,
even though all the fits look similar with each other.

\subsection{Solar reference}
\label{sect:analysis_solar}

\begin{table}
\centering
\caption{Comparison of solar abundances with respect to 
the standard composition of MARCS model atmospheres.}
\label{tab:solar}
\begin{tabular}{lccc} 
\hline
\hline 
Element & Non-LTE  & LTE & \citet{2007SSRv..130..105G}  \\
\hline
Li & $1.00\pm0.04$ & $0.99\pm0.04$ & $1.05\pm0.10$  \\ 
O  & $8.69\pm0.09$  & $8.87\pm0.10$ & $8.66\pm0.05$ \\
Na & $6.16\pm0.03$ & $6.33\pm0.04$ &$6.17\pm0.04$ \\
Mg & $7.62\pm0.02$ & $7.59\pm0.02$ &$7.53\pm0.09$ \\
Al & $6.43\pm0.02$ & $6.46\pm0.02$ &$6.37\pm0.06$ \\
Si & $7.46\pm0.02$ & $7.47\pm0.02$ &$7.51\pm0.04$ \\
Fe & $7.44\pm0.03$ & $7.42\pm0.03$ &$7.45\pm0.05$ \\
\hline
\end{tabular}
\end{table}

In order to obtain accurate abundance ratios of these late-type stars with
respect to the Sun, it is important to determine solar abundances in a
consistent manner \citep[e.g.][]{2006A&amp;A...451..621G}. However, we do
not have access to a high-quality HERMES solar spectrum observed in the
high-resolution mode. Instead, we used the very high-resolution
($R\sim350,000$) Kitt Peak solar flux atlas of
\citet{brault1987spectral}. The solar analysis
proceeded in the same way as for our M67 targets. The resulting
spectroscopic parameters are generally in good agreement with the standard
solar values; the spectroscopic $\teff$~is
lower by $59\,\mathrm{K}$, $\lgg$~is lower by $0.22\,\dex$, and $\feh$~is lower by $0.09\,\dex$,
as we already mentioned in \sect{sect:analysis_parameters}.
The above offsets were applied to the subsequent solar abundance analysis,
as well as to the abundance analysis of all the M67 stars.

We list the final inferred solar abundances in
\tab{tab:solar}. Our solar abundances are in good agreement with those of
\citet{2007SSRv..130..105G}, the most
discrepant elements being magnesium, 
which is $0.09\,\dex$~higher in our non-LTE analysis.
Our solar abundances
are also very similar to
the 1D non-LTE ones presented in \citet{2015A&amp;A...573A..25S, 2015A&amp;A...573A..26S};
all of our values agree with theirs to within $0.04\,\dex$.

\subsection{Error estimation}
\label{sect:analysis_uncertain}

To estimate the overall precision of atmospheric parameters,
we reanalyse all the individual spectra of the 63 stars in our sample that have multiple observations, typically two or three.
We compute the maximum difference in atmospheric parameters obtained from individual spectra and adopt 
the mean values as representative for the entire sample, since we find that these values are 
fairly independent of S/N and stellar parameters.
We finally sum these errors in quadrature with the formal covariance errors returned by \sme~
to obtain the final overall error (effective temperature 
$40\,\mathrm{K}$, surface gravity $0.14\,\dex$ and metallicity $0.07\,\dex$). 

The influence of the uncertainties in the atmospheric parameters on our final abundance determinations
was assessed by varying each time only one of 
atmospheric parameters by the amount of their estimated uncertainties.
We finally added all the individual errors associated with the three contributors quadratically to obtain the total error in abundances. 
These internal errors are applied to produce the error bars in the following \fig{fig:xh_teff}, \fig{fig:lithium},
\fig{fig:xh_teff_atomic} and \fig{fig:xh_logg_atomic}.
Note that the abundance uncertainties may be underestimated, since we have 
not taken into account systematic uncertainties.

\section{Results} 
\label{sect:results}

\begin{figure} 
\includegraphics[scale=1.0,width=\columnwidth]{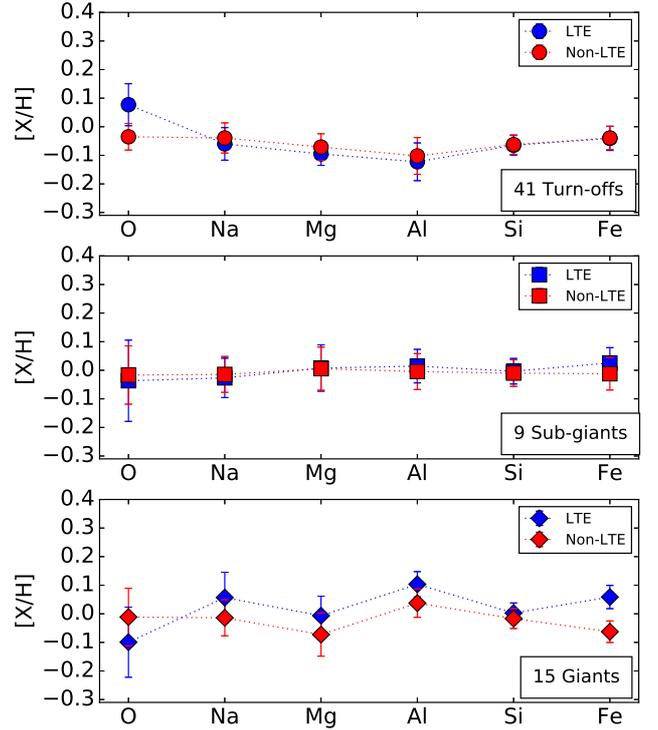}
\caption{Abundance patterns of turn-off, sub-giant and giant stars 
in our final sample.
LTE/non-LTE $\xh{X}$ values were calculated
consistently by treating iron in LTE/non-LTE when
determining the stellar parameters,
and by using our LTE/non-LTE solar reference values.
Each symbol represents the mean abundance 
$\xh{X}$~of that group stars, and
the error bars correspond to the standard deviation in that
group.} 
\label{fig:abundance_pattern}
\end{figure}

\begin{figure*} 
\includegraphics[scale=1.0,width=\textwidth]{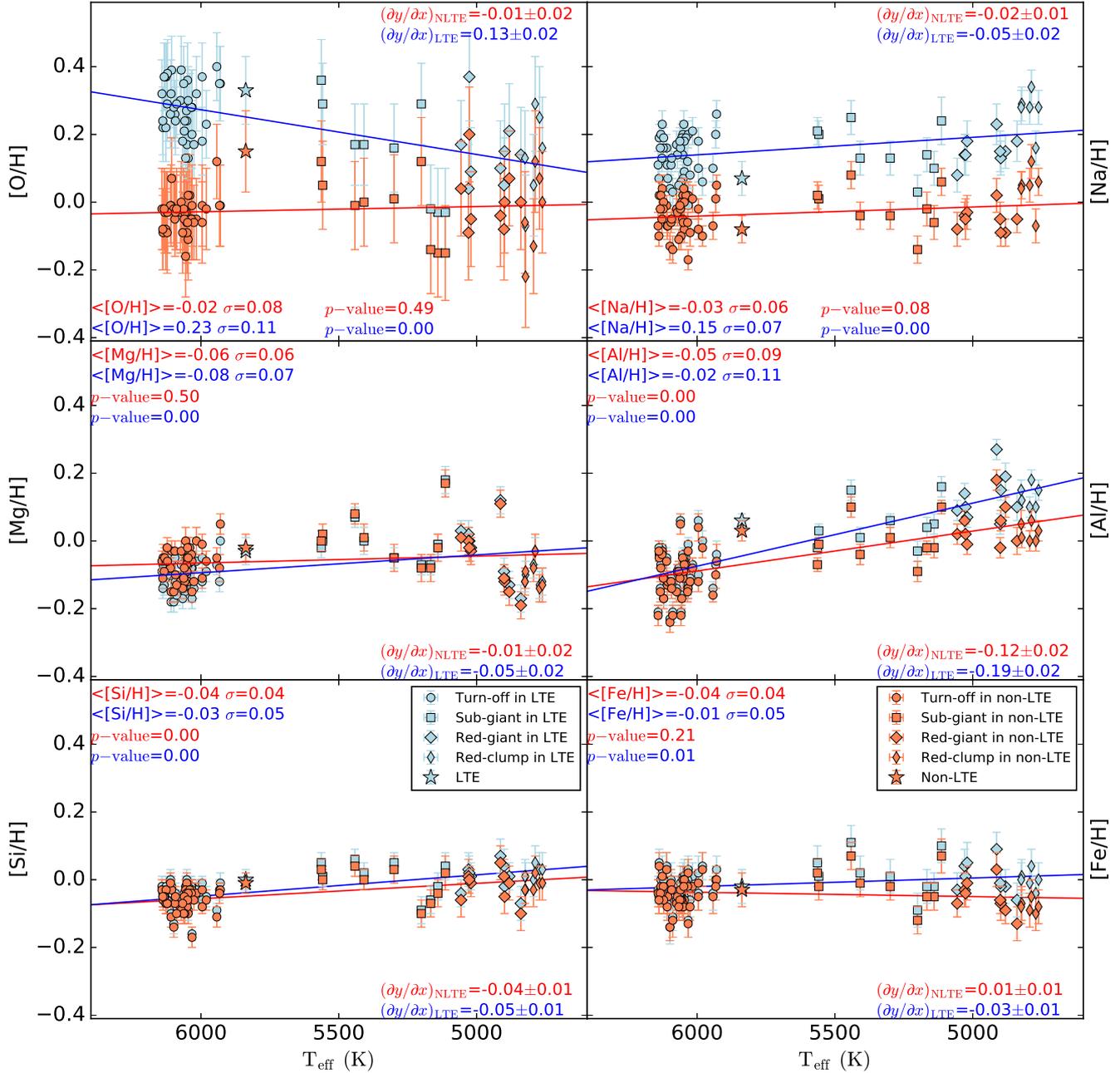} 
\caption{LTE and non-LTE abundances as a function of effective temperature for
individual member stars of M67. All LTE and non-LTE 
abundances shown here were calculated by treating
iron in non-LTE when determining the stellar parameters,
and were put onto a relative 
($\xh{X}$) scale using our non-LTE solar reference.
Stars with different evolutionary states are marked 
using different symbols.
The $p$-values of the trends in LTE and non-LTE 
are shown in the legends, where a small value
(typically $p\text{-value}\lesssim0.05$) is indicative that
the trend is significant with respect to the scatter.
Beyond that, we also list all the gradients (times by 1000) of 
weighted linear fitting lines with the 
standard errors.}
\label{fig:xh_teff}
\end{figure*}

\begin{figure}
\includegraphics[scale=1.0,width=\columnwidth]{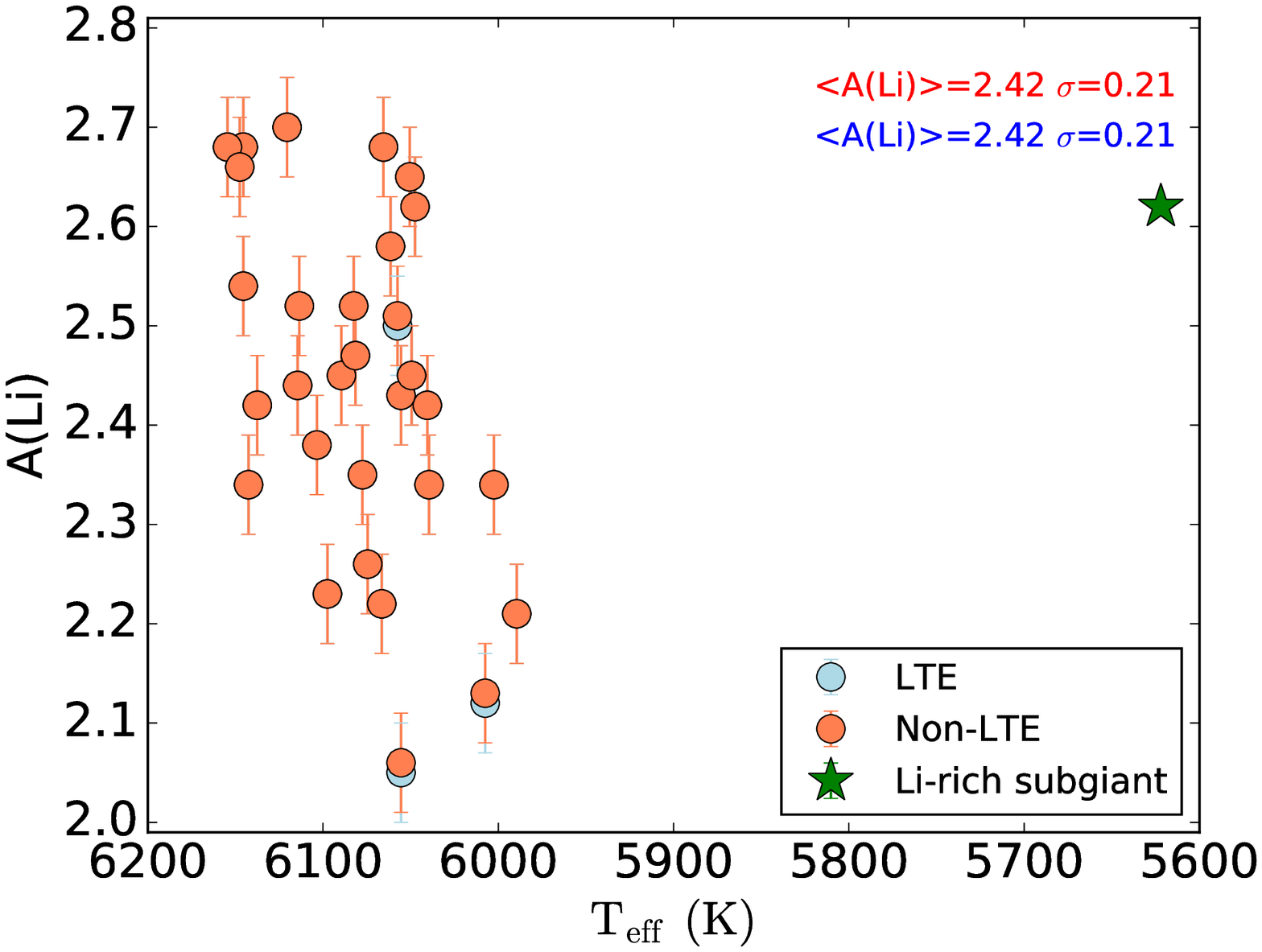}
\caption{Absolute abundance distributions of lithium 
as a function of effective
temperature. A lithium-rich sub-giant located in a binary system,
which we ruled out via our our radial velocity 
criterion, is marked using an asterisk.} 
\label{fig:lithium}
\end{figure}

In order to detail the results of our work,
we first divide our sample into turn-off stars (${\teff}_{\text{; DW}} > 5800\,\mathrm{K}$),
sub-giant stars ($5100\,\mathrm{K} < {\teff}_{\text{; SUB}} < 5800\,\mathrm{K}$), 
and giant stars (${\teff}_{\text{; RGB}} < 5100\,\mathrm{K}$);
in \fig{fig:abundance_pattern} 
we plot the mean $\xh{X}$~abundances for the three groups.
In \fig{fig:xh_teff}~and \fig{fig:lithium}~we plot
LTE and non-LTE abundances of individual M67 stars as a
function of effective temperature.  
We discuss different aspects of these plots in the remainder of this section.

\subsection{Influence of departures from LTE}
\label{sect:results_nlte}

In \fig{fig:abundance_pattern} we compare the mean LTE and non-LTE
$\xh{X}$~abundances for three groups of 
cluster stars: turn-off stars, sub-giant stars, and giant stars.
These were calculated
consistently by treating iron in LTE/non-LTE when
determining the stellar parameters,
and by using our LTE/non-LTE solar reference values. 
Note that part of the absolute NLTE effect on chemical abundances is therefore cancelled and 
only the differential NLTE effects with respect to the Sun are shown in this plot.

For the turn-off stars, under the assumption of LTE, 
we find a large overabundance in $\xh{O}$~of 
more than $0.15\,\dex$, compared to the other species.
This is caused by the non-LTE effect for \ion{O}{I} increasing in magnitude
with increasing effective temperature.
However, under non-LTE,
the abundance ratios $\xh{X}$~for all
elements are generally consistent with each other at slightly sub-solar values.
For the subgiant stars, both LTE and non-LTE abundance results are
generally consistent with each other.
This group also gives results that are closer to the expected 
solar abundances (i.e.~$\xh{X}=0$) than the other two groups.
For the giant stars, the non-LTE abundances
are generally lower than the LTE values, 
and slightly more consistent with a uniform solar composition. 

In \fig{fig:xh_teff} we show
LTE and non-LTE abundances as a function of effective temperature for
individual member stars of M67. Here, both LTE and non-LTE 
abundances were calculated by treating
iron in non-LTE when determining the stellar parameters,
and were put onto a relative 
($\xh{X}$) scale using our non-LTE solar reference.
This illustrates the departures from LTE in the absolute 
abundances, as a function of effective temperature.
We discuss the departures from LTE for different 
elements separately, in the following subsections.

\subsubsection{Lithium}
\label{sect:results_nlte_lithium}

Lithium abundances were determined from the resonance
\ion{Li}{I} $670.8\,\mathrm{nm}$~doublet.  For 
lithium-poor stars ($\mathrm{A(Li)<2}$),
it was impossible to obtain lithium abundances,
because of the very weak line strength.
Most stars cooler than $5900\,\mathrm{K}$~fall into this category,
as they have suffered strong lithium depletion;
an added complication in cooler stars is 
that the doublet is seriously blended with a nearby \ion{Fe}{I} line.
We found one exception at $\teff\approx5600\,\mathrm{K}$,
a lithium-rich sub-giant (\sect{sect:results_li}).
This star was among those that were rejected
as members via the radial velocity criterion.  
The lithium abundances in the
sample are largely insensitive to departures from LTE (see
\fig{fig:lithium}), and the mean Li abundances for non-LTE and LTE
calculations are identical and have the same standard deviation:
$\mathrm{A(Li) = 2.42 \pm 0.21}$. 

The scatter around the mean lithium abundances 
(for those warmer stars in which the doublet 
could be measured) is large ($0.21\,\dex$). This observed spread in our lithium
abundance for stars around the solar mass range has also 
been reported by other M67
studies \citep{2008A&amp;A...489..677P, 2012A&amp;A...541A.150P}.
The fundamental parameters 
of these turn-off stars (mass, metallicity and age) should
be similar; it is possible however that they 
were born with different initial angular momenta,
which is one of the key parameters
for rotational mixing, leading to different
lithium depletions between these otherwise
similar stars \citep{2010IAUS..268..375P}.  
 
All of the turn-offs in the M67 sample in which we detect lithium have
effective temperatures larger than $\teff\approx5900\,\mathrm{K}$; in these
hot turn-off layers, the combination of overpopulation in the Li ground state
and superthermal source function make the non-LTE abundance corrections
approximately zero for this line \citep[e.g.][]{2009A&amp;A...503..541L}.

\subsubsection{Oxygen}
\label{sect:results_nlte_oxygen}

Oxygen abundances were determined from the \ion{O}{I} infra-red triplet,
with its three components
located at $777.19\,\mathrm{nm}$, $777.42\,\mathrm{nm}$, and
$777.54\,\mathrm{nm}$, respectively. The mean non-LTE and LTE abundances of oxygen are
$\xh{O}_{\mathrm{NLTE}}=-0.02\pm0.08$~and
$\xh{O}_{\mathrm{LTE}}=0.23\pm0.11$.  The difference between the oxygen
abundances using non-LTE and LTE synthesis are large
($\mathrm{\Delta_{non-LTE-LTE} \approx -0.25\,\dex}$). The small line
strengths in giant stars and imperfect correction for telluric contamination
result in larger star-to-star scatter compared to the other elements studied
here, even when LTE is relaxed.

The departures from LTE are mainly 
due to photon losses in the lines themselves, 
which leads to an overpopulation of the
metastable lower level, and the increased line opacity strengthens the line
in non-LTE \citep[e.g.][]{1993A&amp;A...275..269K,2003A&amp;A...402..343T,
2016MNRAS.455.3735A}.  As clearly seen in \fig{fig:xh_teff}, the non-LTE
abundance corrections are larger in turn-offs (at higher $\teff$) than in
giants (at lower $\teff$).  This is expected, because the oxygen triplet
gets stronger with effective temperature, increasing the photon losses 
in the lines themselves and
hence making the departures from LTE more severe.

\subsubsection{Sodium}
\label{sect:results_nlte_sodium}

Sodium abundances were determined from the \ion{Na}{I} doublet,
its components located at
$568.26\,\mathrm{nm}$ and $568.82\,\mathrm{nm}$.  Additionally, the
\ion{Na}{I} ($475.18\,\mathrm{nm}$)~line was available for a part of the
sample. The mean non-LTE and LTE abundances of sodium are
$\xh{Na}_{\mathrm{NLTE}}=-0.03\pm0.06$ and
$\xh{Na}_{\mathrm{LTE}}=0.15\pm0.07$. Non-LTE effects evidently play an
important role in Na line formation and cause a substantial negative
correction ($\mathrm{\Delta_{non-LTE-LTE} \approx -0.18\,\dex}$).

The departures from LTE in optical \ion{Na}{I} lines
are largely driven by photon suction in strong
lines, in particular the \ion{Na}{D} resonance lines 
(\ion{Na}{I} $588.9\,\mathrm{nm}$ and 
\ion{Na}{I} $589.5\,\mathrm{nm}$). A
recombination ladder from the \ion{Na}{II} reservoir tends to cause
overpopulations of lower states and subthermal source functions, resulting
in negative abundance corrections that are strongest for saturated lines
\citep[e.g.][]{2011A&amp;A...528A.103L}.

\subsubsection{Magnesium}
\label{sect:results_nlte_magnesium}

Magnesium abundances were determined from three lines;
\ion{Mg}{I} ($473.30\,\mathrm{nm}$),
the \ion{Mg}{I} ($571.11\,\mathrm{nm}$),
and the \ion{Mg}{I} ($769.16\,\mathrm{nm}$).
The mean non-LTE and LTE abundances of magnesium are
$\xh{Mg}_{\mathrm{NLTE}} = -0.06\pm0.06$ and $\xh{Mg}_{\mathrm{LTE}}=
-0.08\pm0.07$.  Although the impact of departures from LTE is not very
pronounced on the mean abundances, 
it is interesting to note there is still a clear influence on
the abundance trends. 
This is because the giants tend to have negative abundance
corrections, whereas the turn-offs
tend to have positive abundance corrections.

The physical non-LTE effect is different in turn-offs 
and giants. In turn-off stars, 
the photoionisation rates for the lower \ion{Mg}{I}~levels are
substantial, which can lead to overionisation,
resulting in positive non-LTE abundance corrections.
In contrast, in giant
stars, \ion{Mg}{I} lines
(especially the \ion{Mg}{I} $571.11\,\mathrm{nm}$~line)
suffer from photon losses, making the 
abundance corrections negative
\citep[e.g.][]{2015A&amp;A...579A..53O, 2017ApJ...847...15B}.

\subsubsection{Aluminum}
\label{sect:results_nlte_aluminium}

Aluminium abundances were determined using the doublet:
\ion{Al}{I} ($669.6\,\mathrm{nm}$)~and \ion{Al}{I} ($669.8\,\mathrm{nm}$).
The mean non-LTE and LTE abundances of aluminium are
$\xh{Al}_{\mathrm{NLTE}}=-0.05\pm0.09$~and 
$\xh{Al}_{\mathrm{LTE}}=-0.02\pm 0.11$.
The very weak aluminium lines in turn-offs cause a
substantial abundance scatter.  
In addition, the doublet
falls in a spectral region where the wavelength calibration
of HERMES is of lower quality, which manifests itself in poor synthetic fits
to the observed spectral lines.  To improve this defect, we set radial
velocity as a free parameters when carrying out spectra synthesis of
aluminium; this unfortunately further contributes to the abundance scatter.

The non-LTE abundance correction are always
negative and become much
more severe in giants than the corrections in turn-offs.
The negative sign of the corrections is due to photon suction effects,
resulting in overpopulations of lower levels and subthermal source
functions. These effects are strongest in giants.
Towards warmer effective temperatures, 
the non-LTE effect starts to change:
a larger supra-thermal UV radiation field
means that a competing overionisation effect becomes more efficient.
As such, the non-LTE abundance corrections are much less
less severe in turn-offs \citep{2017arXiv170801949N} .

\subsubsection{Silicon}
\label{sect:results_nlte_silicon}

Five silicon lines were used to determine silicon abundances:
\ion{Si}{I} ($566.55\,\mathrm{nm}$); 
\ion{Si}{I} ($569.04\,\mathrm{nm}$);
\ion{Si}{I} ($570.11\,\mathrm{nm}$); 
\ion{Si}{I} ($579.31\,\mathrm{nm}$), 
and \ion{Si}{I} ($672.18\,\mathrm{nm}$).
The mean non-LTE and LTE abundances of silicon are
$\xh{Si}_{\mathrm{NLTE}}=-0.04\pm0.04$~and 
$\xh{Si}_{\mathrm{LTE}}=-0.03\pm 0.05$.  

The non-LTE abundance corrections for Si lines are
not very pronounced, however they are always negative in this sample.
Generally, photon losses in the \ion{Si}{I} lines drives overpopulation for
the lower levels and underpopulation for higher levels,  which strengthen
the lines in non-LTE.

\subsubsection{Iron}
\label{sect:results_nlte_iron}

Iron abundances were determined from 
a selection of \ion{Fe}{I} and \ion{Fe}{II} lines,
that are listed in \cite{2018MNRAS.478.4513B}.
The mean non-LTE and LTE
abundances of iron are $\xh{Fe}_{\mathrm{NLTE}}=-0.04\pm0.04$ and
$\xh{Fe}_{\mathrm{LTE}}=-0.01\pm0.05$. Non-LTE effects cause a small
negative correction ($\mathrm{\Delta_{non-LTE-LTE} \approx -0.03\,\dex}$).

Since \ion{Fe}{II} lines are almost immune to non-LTE effects in late-type
stars \citep[at least, in 1D hydrostatic model atmospheres
such as those used in this work --
in 3D hydrodynamic model atmospheres this is not always the case;
e.g.][Table 3]{2016MNRAS.463.1518A}, the main contribution 
to the difference between the mean 
abundances under the assumption of LTE and non-LTE 
comes from the \ion{Fe}{I} lines. The traditional
non-LTE effect for \ion{Fe}{I} lines is overionisation;
at solar-metallicity, however, this effect is relatively small,
and photon losses in the \ion{Fe}{I} lines as well as a general
photon-suction effect are more important.
We therefore see slightly negative abundance corrections.
The effects are more severe in giants, where
these intermediate-excitation \ion{Fe}{I} lines are stronger.

\subsection{Lithium-rich sub-giant}
\label{sect:results_li}

Among the full sample of stars observed in the M67 field, we discovered a
sub-giant star (S95) with a very high lithium abundance $\mathrm{A(Li)=2.6}$~(see
\fig{fig:lithium}).  However, because of its radial velocity,
$\mathrm{RV}=38.5\,\kms$, which is high compared to the cluster mean
(see \fig{fig:radv_metallicity}), we regard this star as a potential
non-member and have excluded it from the discussion of cluster abundance
trends.  No other sub-giant star in the sample has such a high lithium
abundance, and severe lithium depletion is expected at this evolutionary
stage after leaving the main sequence turn-off
\citep{1995ApJ...446..203B,2012A&amp;A...541A.150P}.  By checking the
position and magnitude information, this star has been confirmed as a
spectroscopic binary in the SIMBAD.

\cite{2006A&amp;A...451..993C} also reported a
lithium-rich sub-giant star S1242 with ($\mathrm{A(Li)=2.7}$).
S1242 has been 
verified as a member of a large eccentricity binary system in M67, with a
faint low-mass dwarf companion providing negligible contribution to the
luminosity \citep{1977A&amp;AS...27...89S, 1990AJ....100.1859M}.
\cite{2006A&amp;A...451..993C} proposed that high chromospheric activity 
and unusually high rotational velocity of S1242 may be induced by tidal interaction, 
which could help the star conserve its lithium abundance from the turn-off stage.
Interestingly, \cite{2014A&amp;A...562A.102O} also found a lithium-rich sub-giant star 
S1320 with $\mathrm{A(Li)=2.3}$.
This sub-giant has been included in their membership, since they did not find any 
evidence that this star has been contaminated by a companion. 
It is worth to follow up these stars, as the identification of these stars should 
prove useful for providing insight into 
the processes in binaries that can affect the surface abundances.


\subsection{Abundance trends}
\label{sect:results_trends}

As illustrated in \fig{fig:xh_teff}, we have found abundance
trends with effective temperature for some elements.
The trends are more pronounced when LTE
is assumed; furthermore, the scatter around the mean for oxygen
becomes more pronounced when LTE is assumed.
Even under the assumption of non-LTE, however,
there still exist some systematic abundance
differences between turn-offs, sub-giants and giants,
as can be seen in \fig{fig:xh_teff}.

To determine if there is a significant correlation between 
element abundance and effective temperature, 
we calculate $p$-values in the linear regression analysis by
assuming there is no correlation between these two parameters
in the null hypothesis.
The $p$-values of the trends are shown in the legends 
of \fig{fig:xh_teff}, where a small value
(typically $p\text{-value}\lesssim0.05$) is indicative that
the trend is significant with respect to the scatter.
We can thus say that, under the assumption of LTE, the trends in surface abundance
against effective temperature are significant with respect to the scatter,
for all of the species shown in \fig{fig:xh_teff}.
In contrast, under the assumption of non-LTE, the trends for oxygen, 
sodium, magnesium and iron are not significant with respect to the scatter, 
while for aluminium and silicon the trends remain significant.
We further note an obvious deviation from the linear trend in the 
behaviour of Mg abundance with effective temperature; subgiants appear 
overabundant with respect to the linear trend and red giants underabundant.

In summary, non-LTE analysis tends to flatten the trends with effective
temperature seen in LTE, which reduces the scatter in mean abundance for
all the elements, when the full sample is considered. The remaining residual trends may reflect other systematic
errors still present in the analysis or be intrinsic to the cluster. We shall consider this in more detail in
\sect{sect:discussion}.

\section{Discussion}
\label{sect:discussion}

\subsection{Comparison with atomic diffusion models}
\label{sect:discussion_models}

\begin{figure*} 
\includegraphics[scale=1.0,width=\textwidth]{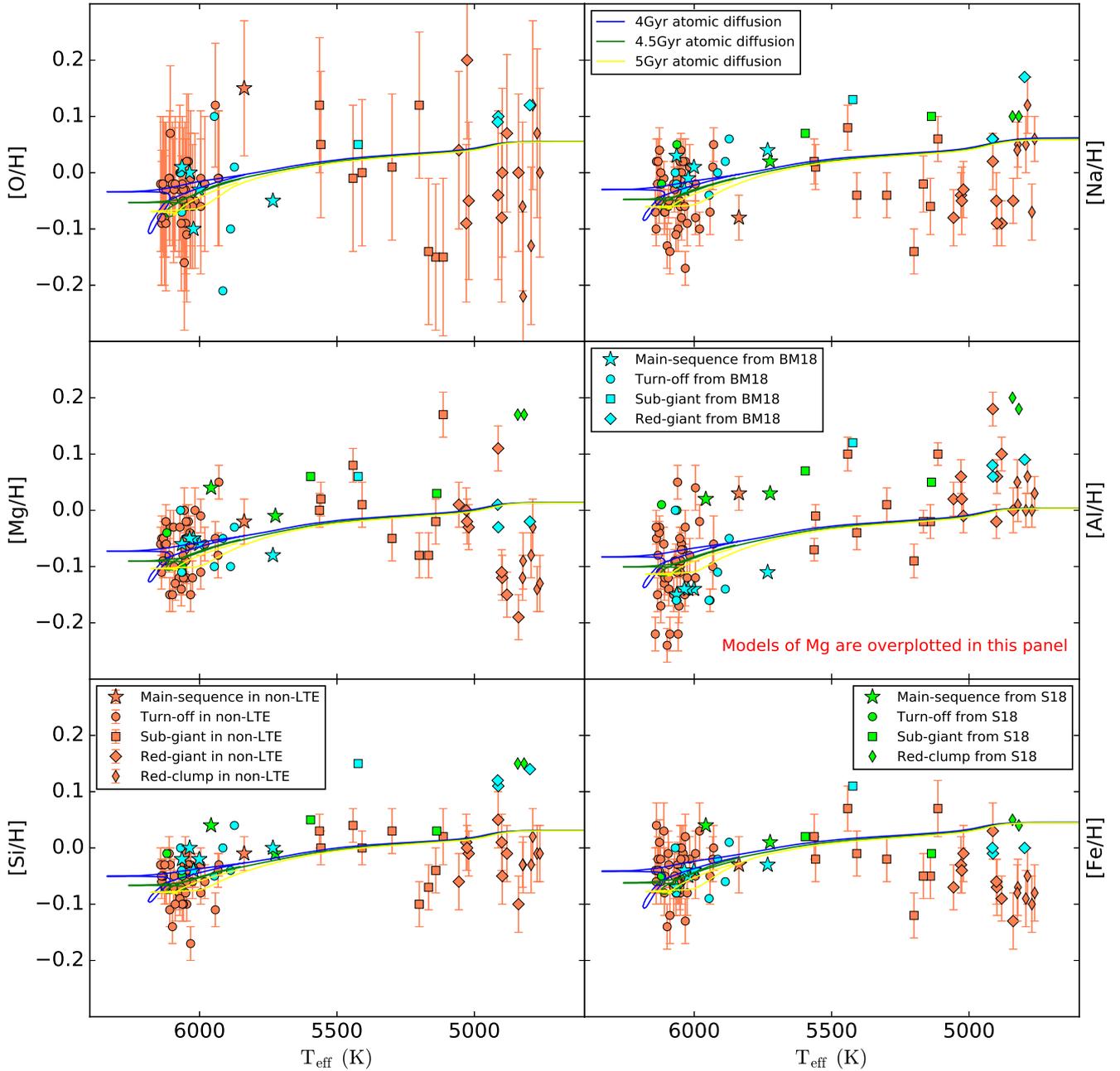} 
\caption{Non-LTE Abundances 
$\xh{X}$~as a function of effective temperature for
individual M67 stars. We overplot surface abundance isochrones 
from atomic diffusion models with solar metallicity
and different evolution ages. Al is not shown in the model-data comparison,
since it has been neglected in the model output. Instead, we overplot the models of Mg
on the Al measurements.
We also overplot the abundance
results from \citet{2018ApJ...857...14S} and \citet{2018MNRAS.478..425B}. 
Stars in different evolutionary states are marked
with different symbols.}
\label{fig:xh_teff_atomic}
\end{figure*}

\begin{figure*} 
\includegraphics[scale=1.0,width=\textwidth]{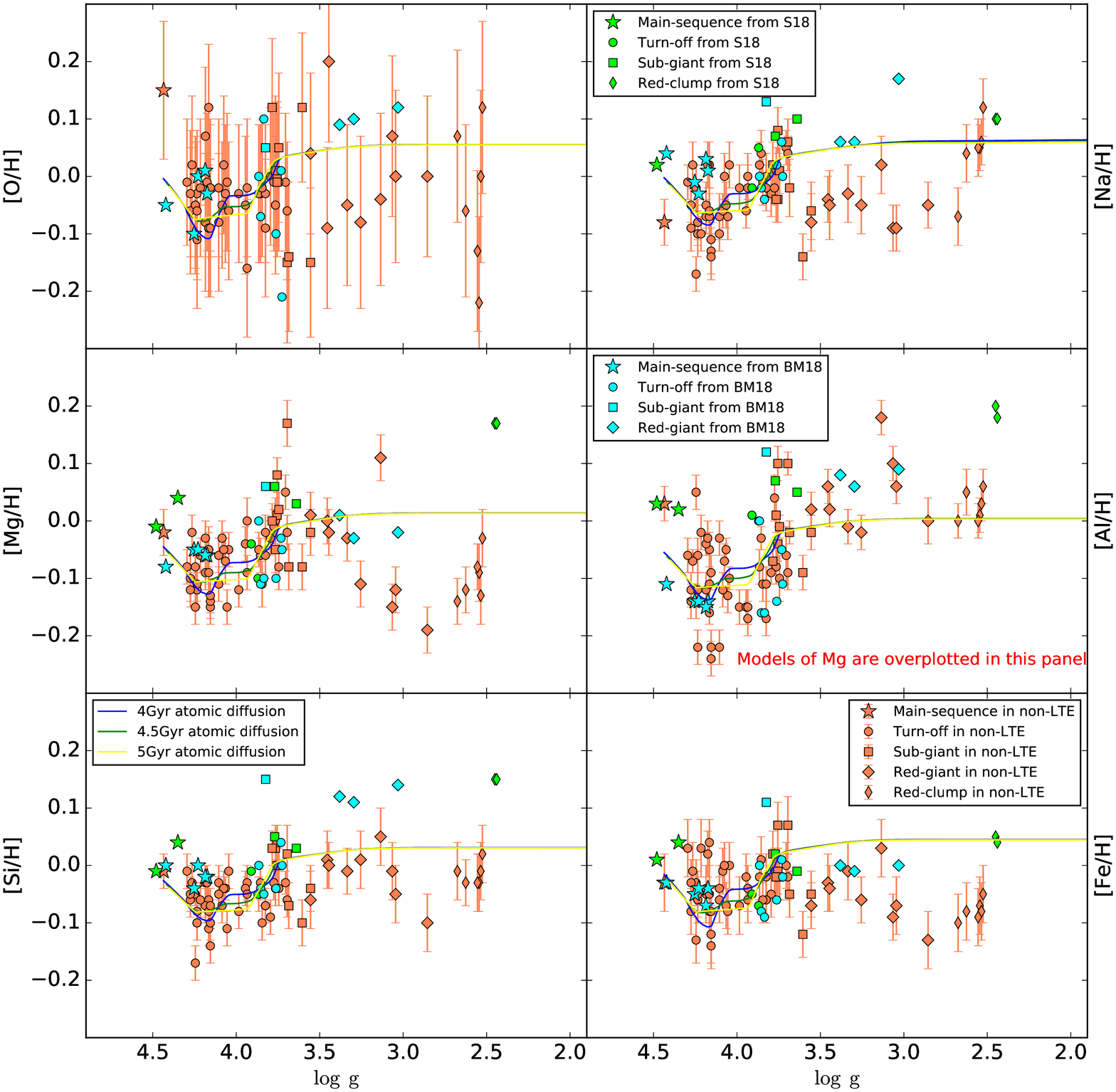} 
\caption{Non-LTE Abundances 
$\xh{X}$~as a function of $\lgg$ for
individual M67 stars. We overplot surface abundance isochrones 
from atomic diffusion models with solar metallicity
and different evolution ages. Al is not shown in the model-data comparison,
since it has been neglected in the model output. 
Instead, we overplot the models of Mg on the Al measurements.
We also overplot the abundance
results from \citet{2018ApJ...857...14S} and \citet{2018MNRAS.478..425B}.
Stars in different evolutionary states are marked
with different symbols.}
\label{fig:xh_logg_atomic}
\end{figure*}

Atomic diffusion is a continuous process whose influence immediately below
the outer convection zone causes surface abundance variations during the
main-sequence phase of a star.  At the turn-off point, where the convective
envelope is the thinnest, the settling of elements reaches a maximum.  As
the star evolves along the sub-giant branch and red giant branch, the surface
abundances begin to recover gradually to the initial value due to the
enlarged surface convection zone, except for those light elements that are
affected by nuclear processing.

The metals in our Sun are thought to be underabundant relative to the
initial bulk composition, by about
$0.04\,\dex$~\citep[e.g.][]{2009ARA&amp;A..47..481A}.
\citet{1998ApJ...504..539T} demonstrated that the diffusive process is
dominant at the end of the main-sequence phases of solar-type stars, thus the
turn-off stars in M67 with comparable age to the Sun may show even larger
effects of atomic diffusion.  Larger effects are also expected in warm
metal-poor stars, because of their 
older ages and thinner surface convection
zones \citep{1984ApJ...282..206M}.

Our sample includes stars in different evolutionary
states, including main-sequence, turn-off, sub-giant,
red-giant and red-clump stars.
It is therefore of interest to compare our 
results with those predicted by stellar evolutionary models
that include atomic diffusion.
We adopted the surface abundances that were calculated
in \citet{2017ApJ...840...99D} with solar metallicity, 
initial masses ranging from $0.5 \rm{M}_{\odot}$ to $1.5 \rm{M}_{\odot}$ 
and ages of $t=4.0\,\Gyr$, $t=4.5\,\Gyr$ and $t=5.0\,\Gyr$, respectively.
The stellar evolutionary models 
(\mist;~\citealt{2016ApJS..222....8D,2016ApJ...823..102C})
have included atomic diffusion, overshooting mixing and turbulent diffusion.
Furthermore, the models are calculated with radiative acceleration, which 
acts differently on different chemical species and can thus
potentially explain different abundance trends for 
the different elements under consideration.

In \fig{fig:xh_teff_atomic} and \fig{fig:xh_logg_atomic}, we overplot the stellar evolutionary models 
on our results for the surface abundances versus effective temperature and gravity, respectively.
Since Al has been neglected in the model output,
models of Al are not shown in the model-data comparison. We note that Al is expected 
to behave similarly to the other elements \citep[see e.g.][]{2018MNRAS.478..425B}.
We thus overplot the models of Mg on the Al measurements instead.
Since the zero-points of the models are not relevant here, and
we are more interested in the effect of atomic diffusion on their relative surface abundances,
small arbitrary offsets have been applied to all the model abundances so as to generally
match our abundance measurements for the turn-off stars.

\fig{fig:xh_logg_atomic} most clearly illustrates the evolutionary effects predicted by the 
models; the model abundances decrease on the main-sequence with increasing mass
to reach a minimum around the turn-off; 
the severity of this depletion is age-dependent, being more severe for older ages.
Moving to later evolutionary stages (lower surface gravity and effective temperature),
the elements are brought back up to the surface by convective mixing 
(i.e.~the first dredge-up), and the surface abundance depletion becomes less severe.
At the base of the red giant branch, the surface abundances are restored to 
the original composition; the models actually predict a slight increase in the surface abundances 
over the initial values as a result of hydrogen being consumed during central H-burning.

We now highlight some interesting aspects evident from the comparison 
between our observed abundances and the model predictions in \fig{fig:xh_logg_atomic}. We note that the initial decrease
with increasing mass cannot be tested with our data, since there are too few main sequence stars.
However, there is a satisfying morphological agreement with the models in the dredge-up pattern from the
turn-off to the subgiant branch. However, our abundance measurements of the red-giant and red-clump stars 
(with effective temperature less than $5100\,\mathrm{K}$) do not fit the predicted trend very well, 
even considering the abundance errors, for all elements except possibly Al. One possible reason for this discrepancy could be 
that the stellar parameters for these giant stars are poorly determined (see \fig{fig:isochrone}). 
Problems in the main stellar parameters will propagate the systematic
offsets to the individual stellar abundances. However, this can not be the single contribution to explain this discrepancy, 
since the systematic offsets propagating to different elements may have different correction directions.

\subsection{Comparison to other studies}
\label{sect:discussion_comparison} 

\begin{table*}
\begin{center}
\caption{The comparison of average abundances in common for M67 based on
high resolution spectroscopy. The total number of stars analyzed in each study is given by $\#$.}
\label{tab:comparison}
\begin{tabular}{lcccccccccc}
\hline
\hline
&
& $\#$
& R
& SNR
& $\feh$
& $\xfe{O}$
& $\xfe{Na}$ 
& $\xfe{Mg}$
& $\xfe{Al}$
& $\xfe{Si}$  \\
\hline
& NLTE$^{1}$
& $66$
& $42000$ 
& $50$--$150$
& $-0.04\pm0.04$
& $+0.04\pm0.09$ 
& $+0.03\pm0.05$ 
& $+0.00\pm0.05$
& $+0.01\pm0.07$
& $+0.02\pm0.03$ \\
& T00$^{2}$
& $9$
& $30000$--$60000$
& $\ge100$
& $-0.03\pm0.03$
& $+0.02\pm0.06$
& $+0.19\pm0.06$  
& $+0.10\pm0.04$  
& $+0.14\pm0.04$ 
& $+0.10\pm0.05$  \\
& Y05$^{3}$
& $3$
& $28000$
& $30$--$100$
& $+0.02\pm0.14$
& $+0.07\pm0.05$ 
& $+0.30\pm0.10$   
& $+0.16\pm0.08$ 
& $+0.17\pm0.05$
& $+0.09\pm0.11$ \\
& R06$^{4}$
& $10$
& $45000$
& $90$--$180$
& $+0.03\pm0.03$
& $+0.01\pm0.03$ 
& $+0.05\pm0.07$
& $+0.00\pm0.02$  
& $-0.05\pm0.04$ 
& $+0.02\pm0.04$ \\
& P08$^{5}$ 
& $6$
& $100000$
& $\simeq80$
& $+0.03\pm0.04$
& $-0.07\pm0.09$ 
& $-0.02\pm0.07$
& -        
& $-0.03\pm0.11$ 
& $-0.03\pm0.06$  \\
& P10$^{6}$ 
& $3$
& $30000$
& $50$--$100$
& $+0.05\pm0.02$   
& $+0.04\pm0.10$ 
& $+0.08\pm0.09$  
& $+0.27\pm0.04$
& $+0.03\pm0.02$  
& $+0.10\pm0.02$ \\
& F10$^{7}$
& $3$
& $30000$ 
& $150$--$180$
& $+0.03\pm0.07$ 
& $-0.16\pm0.05$
& $+0.13\pm0.10$   
& $+0.05\pm0.03$ 
& $+0.11\pm0.07$ 
& $+0.18\pm0.04$ \\ 
& {\"O}14$^{8}$  
& $14$
& $50000$ 
& $150$ 
& $-0.02\pm0.04$  
& $-0.02\pm0.05$  
& $+0.02\pm0.03$ 
& $+0.02\pm0.02$ 
& $+0.02\pm0.04$ 
& $-0.01\pm0.02$ \\
& BM18$^{9}$  
& $15$ 
& $47000$ 
& $-$ 
& $-0.02\pm0.05$ 
& $+0.02\pm0.09$ 
& $+0.06\pm0.04$ 
& $-0.02\pm0.02$
& $-0.04\pm0.07$
& $+0.05\pm0.04$ \\
& S18$^{10}$ 
& $8$ 
& $22500$ 
& $120$--$956$ 
& $+0.00\pm0.04$ 
& $-$ 
& $+0.06\pm0.04$ 
& $+0.04\pm0.06$ 
& $+0.07\pm0.05$ 
& $+0.04\pm0.04$ \\
\hline 
\end{tabular} 
\medskip
\small \item \textbf{Notes.} 
(1) This work
(2) \citet{2000A&amp;A...360..499T} analysed 6 red-clump stars and 3
giant stars. (3) \citet{2005AJ....130..597Y} analysed 3 red-clump
stars. (4) \citet{2006A&amp;A...450..557R} analysed 8 dwarfs and 2
slightly evolved stars. (5) \citet{2008A&amp;A...489..403P} analysed 6
main-sequence stars. (6) \citet{2010A&amp;A...511A..56P} analysed 3 red-clump stars. 
(7) \citet{2010AJ....139.1942F} analysed 3 red-clump stars. 
(8) \citet{2014A&amp;A...562A.102O} analysed 14 stars whose 6 are
located on the main-sequence, 3 are at the turn-off point, and 5 are
on the early sub-giant branch.
(9) \citet{2018MNRAS.478..425B} analysed 15 stars whose 5 are located on the main-sequence,
6 are at the turn-off phase, 1 are on the sub-giant branch and 3 are on the red-giant branch.
(10) \citet{2018ApJ...857...14S} analysed 8 stars, including two main-sequence stars, two turn-off stars,
two sub-giants and two red-clump stars
\end{center}
\end{table*}

\begin{figure} 
\includegraphics[scale=1.0,width=\columnwidth]{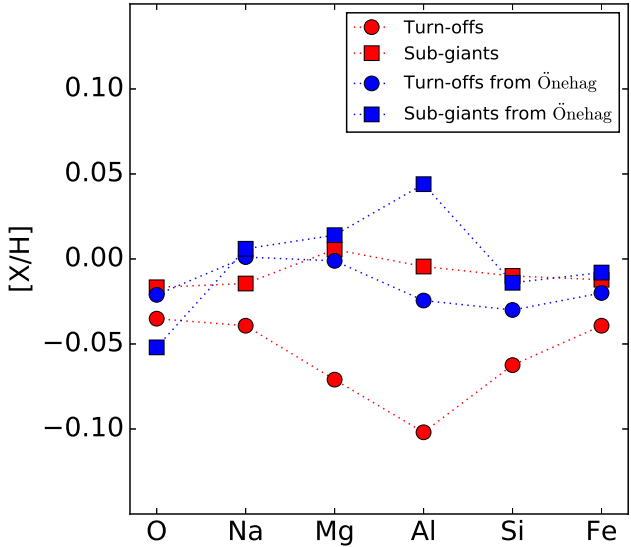} 
\caption{A comparison between our non-LTE abundance patterns of turn-off, sub-giant and giant stars
and those from \"Onehag's turn-off and early sub-giant stars.}
\label{fig:xh_pattern_onehag}
\end{figure}

In this section, we compare our abundance results to previous
high-resolution studies of M67. 
\tab{tab:comparison}~summarises the target
selection and spectroscopic quality for seven literature studies. We also
include the mean abundance ratios determined in those studies.
We compare these results, which were mainly based on
equivalent-widths and under the assumption of LTE,
with our own results, which are based on spectral line fitting
and under non-LTE.

Our mean [Fe/H] value in non-LTE for M67 is consistent with the value of
\citet{2000A&amp;A...360..499T}, \citet{2014A&amp;A...562A.102O} and \citet{2018MNRAS.478..425B}, but
is slightly lower than those determined from 
the other studies
shown in \tab{tab:comparison}. Generally
all the results are comparable 
with solar metallicity to within their respective errors.
However, some disagreements between other measured abundances from different
studies do exist.

Overall, our abundance ratios in non-LTE are close to solar, and are
systematically lower than those studies 
wherein only giants have been analyzed,
namely \citet{2000A&amp;A...360..499T},
\citet{2005AJ....130..597Y}, \citet{2010A&amp;A...511A..56P} and
\citet{2010AJ....139.1942F}. 
The abundance results that are mainly based on unevolved stars from
\citet{2006A&amp;A...450..557R}, \citet{2008A&amp;A...489..403P}, 
\citet{2014A&amp;A...562A.102O},  \citet{2018MNRAS.478..425B} and \citet{2018ApJ...857...14S}
are more consistent with those presented 
in this work. 

The differences in the abundances determined in this work and those presented
elsewhere could be the result of a variety of factors, including
the choice of atmospheric model, abundance calculation code, 
the determined stellar parameters, the choice of $\log gf$ values and line lists, 
the choice of solar reference abundances and non-LTE effects.
In this work, all of the abundances are determined by spectrum synthesis, 
which are more reliable and accurate, especially when the lines are 
blended, than the traditional equivalent width analysis.
We note, too, that our results benefit from being based on the largest sample of 
high-quality spectra yet published, 
covering turn-off, sub-giant star, red giant and red clump stars 
compared with other studies, whose abundances are derived based on a smaller number of objects. 

We compare the results 
of \citet{2014A&amp;A...562A.102O}
with those presented in this work in \fig{fig:xh_pattern_onehag}.
\citet{2014A&amp;A...562A.102O} analysed $14$~turn-offs and sub-giants
using high resolution spectra ($R\approx50,000$),
an analysis based on equivalent-widths and under the assumption of LTE.
Their abundances were derived for each spectral line 
individually relative to those of the solar proxy M67-1194.
Our mean chemical abundances are typically lower
than the ones from \citet{2014A&amp;A...562A.102O}.
However, in that work as well as our own,
we find that the abundances in sub-giants
are enhanced relative to those in turn-offs.
This enhancement is smaller in the results of
\citet{2014A&amp;A...562A.102O} than in this work;
this may be because the sub-giants used in that work
are located very close to the turn-off,
whereas here they span the full subgiant branch.
These overall increasing abundances from turn-offs
to sub-giants could be a signature for possible diffusion process
(\sect{sect:discussion_models}).

Recent studies by \cite{2018MNRAS.478..425B} and 
\cite{2018ApJ...857...14S} both investigated the presence of atomic diffusion effects in M67
by analysing the member stars across different evolutionary phases. We overplot
their results in \fig{fig:xh_teff_atomic} and \fig{fig:xh_logg_atomic}.
Their inferred abundance patterns show an overall agreement with the atomic diffusion models from 
\citet{2017ApJ...840...99D} and their abundance distributions for turn-off and subgiant stars are 
generally consistent with our non-LTE results, with some notable exceptions. We note that 
the other two studies show no evidence of low abundances for red giants compared to less 
evolved stars, as seen in our data for O, Na, Mg, and Fe. This reinforces our suspicion that our giant star abundances 
are not accurate (see \sect{sect:discussion_models}) 

Looking at individual elements, the measured [O/H] from \cite{2018MNRAS.478..425B} also presents a fairly large scatter, which the authors ascribe to telluric blending and weakness of the [OI] line at 630nm. However, 
this line is not expected to suffer large non-LTE effects and the agreement with our non-LTE abundances
is significantly better than with our LTE abundances. The LTE [Na/H] abundances derived by the other two groups 
are consistently somewhat higher than our non-LTE abundance trend and \cite{2018MNRAS.478..425B}  
estimate that their Na LTE abundances are indeed overestimated by $0.1-0.15\,\dex$. 

The [Mg/H] abundances agree well for unevolved stars, while the red giants show a disagreement of $>0.2\,\dex$
between the three groups, which cannot be attributed to non-LTE effects. 
For [Al/H], our abundances tend to fall between results of the other two groups, but there is 
satisfactory agreement on the increasing abundance trend with evolutionary phase.
The [Si/H] abundances of the other two groups are higher than ours and the predicted abundance trend 
slightly steeper. We note that \cite{2018MNRAS.478..425B} suspect that their Si analysis suffers
from an unknown bias, elevating the abundances in giants with respect to dwarfs. 
The [Fe/H] abundances are in good agreement between all three studies for turn-off stars and subgiants, but not for giants,
as mentioned above.

\section{Conclusion} 
\label{sect:conclusion}

We have presented a comprehensive determination of the M67 elemental
abundances of lithium, oxygen, sodium, magnesium, aluminium, silicon, and
iron.  We analysed lines using non-LTE and LTE calculations with 1D
hydrostatic \marcs~model atmospheres based on high resolution, high quality
spectra from the GALAH survey.

We have accounted for non-LTE effects in the line formation of different
elements.  For lithium, non-LTE effects are not prominent. 
However, the large scatter ($0.21\,\dex$) in lithium abundances in stars with
similar stellar parameters (i.e.~mass, metallicity and age) may indicate that the
stars in this cluster could have different initial angular momentums to
which would naturally result in different levels of lithium depletion.
In addition, we found a lithium-rich
sub-giant in our sample, which we 
note is a spectroscopic binary. 
It could be a potential candidate to
study unusual lithium induced by tidal effects.

We found that the scatter in mean abundance is reduced for all the 
elements under the assumption of non-LTE, compared to under LTE, 
because non-LTE analyses flattens
the trends in surface abundances with
effective temperature (see \fig{fig:xh_teff}).  However, abundance
differences between stars in different evolutionary phases are not fully
erased by non-LTE effects. The star-to-star abundance scatter for 
similar stars appears largely unaffected by non-LTE analysis.

We compared our observed abundance trends
with the trends predicted by the atomic diffusion model of
\citet{2017ApJ...840...99D}, assuming solar metallicity and approximately
solar age. Our non-LTE results match well with model prediction 
for turn-off stars and subgiants within the errors, however, they fail
to meet the predicted trend for later phase red-giant and red-clump stars. 
One possible reason for this differences could be caused by the poor determination of 
stellar parameters for those giant stars. 

To increase the accuracy of our abundance measurements further,  3D hydrodynamical model
atmospheres should be considered. Such modelling is important for late type
atmospheres, where the spectral line form at the top of the convective
region, and eliminates the need for the artificial broadening parameters,
such as microturbulence and macroturbulence.
\citep[e.g.][]{2000A&amp;A...359..729A}. Performing a 3D non-LTE study is 
beyond the scope of the present work.  We
note however that 3D corrections for the same lines can go in opposite
directions for turn-offs and giants.  Consequently, it is 
possible that a 3D non-LTE analysis would find significantly flatter or steeper
abundance trends than those presented in \sect{sect:results_trends} \citep{2007ApJ...671..402K}.

Finally, we underline the necessity to include accurate non-LTE corrections
in order to obtain more reliable abundances to study abundance evolution and
chemical tagging.  Our analysis shows that, due to the potential influence
of both systematic abundance errors and of stellar evolution effects, the
method of connecting stars in the field to a common birth location by
chemical similarity is significantly more reliable for stars in the same
evolutionary phase.

\section*{Acknowledgements}
\label{acknowledgements}
XDG, KL, and AMA acknowledge funds from the Alexander von Humboldt Foundation in
the framework of the Sofja Kovalevskaja Award endowed by the Federal Ministry of
Education and Research, and KL also
acknowledges funds from the Swedish Research Council 
(grant 2015-004153) and Marie Sk{\l}odowska Curie Actions 
(cofund project INCA 600398). TZ acknowledges financial support of the Slovenian Research Agency 
(research core funding No. P1-0188).
SLM acknowledges support from the Australian Research Council through grant DE140100598.
Parts of this research were conducted by the Australian Research Council Centre of Excellence for All Sky Astrophysics in 3 Dimensions (ASTRO 3D), through project number CE170100013. 
This work is also based on data acquired from the Australian Astronomical Telescope. 
We acknowledge the traditional owners of the land on which the AAT stands, the Gamilaraay people, and pay our respects to elders past and present.
This research has made use of the SIMBAD database,
operated at CDS, Strasbourg, France.
We thank an anonymous referee for many insightful comments that helped improve the manuscript.

\bibliographystyle{mnras}
\bibliography{astronomy_ref}
\begin{appendix}
\section{Tables of the fundamental parameters}
\begin{table*}
\begin{center}
\caption{Fundamental parameters of the sample stars from the spectroscopic analysis of GALAH data. The columes from left to right show the GALAH ID, the star identifier in the 2MASS catalogue, the type of the star,the effective temperature, the surface gravity, the stellar metallicity, the micro-turbulence, the projected surface rotational velocity and radial velocity. Note that $\vsini$ is actually a combined measurement from both $\vsini$ and $\vmac$, since they have a degenerate influence effect on spectral line broadening and cannot been disentangled.}
\label{tab:fund-paras}
\begin{tabular}{c c c c c c c c c}
\hline
$\text{GALAH ID}$ &
$\text{2MASS ID}$ &
$\text{Group}$ &
$\teff$\,($\mathrm{K}$) &
$\lgg$ &
$\text{Metallicity}$ &
$\vmic$\,($\kms$) &
$\vsini$\,($\kms$) &
$\text{RV}$\,($\kms$) \\
\hline
\hline
$6561552$ & $08505344+1144346$ & $\text{Main-sequence}$ & $5837$ & $4.43$ & $-0.05$ & $0.93$ & $8.81$ & $34.65$ \\
$6560101$ & $08511833+1143251$ & $\text{Turn-off}$ & $6141$ & $4.11$ & $-0.07$ & $1.07$ & $7.70$ & $34.06$ \\
$6577714$ & $08514522+1156552$ & $\text{Turn-off}$ & $6138$ & $4.11$ & $-0.08$ & $1.09$ & $8.30$ & $34.62$ \\
$6554484$ & $08514493+1138589$ & $\text{Turn-off}$ & $6137$ & $4.16$ & $0.01$ & $1.08$ & $7.68$ & $33.36$ \\
$6577148$ & $08505439+1156290$ & $\text{Turn-off}$ & $6133$ & $3.93$ & $-0.08$ & $1.08$ & $9.20$ & $33.97$ \\
$6569011$ & $08511534+1150143$ & $\text{Turn-off}$ & $6131$ & $3.87$ & $-0.05$ & $1.08$ & $9.15$ & $34.14$ \\
$6565966$ & $08504766+1147525$ & $\text{Turn-off}$ & $6127$ & $4.18$ & $-0.06$ & $1.08$ & $7.76$ & $34.85$ \\
$6565326$ & $08511476+1147238$ & $\text{Turn-off}$ & $6126$ & $3.86$ & $-0.03$ & $1.08$ & $9.54$ & $35.38$ \\
$6574584$ & $08514122+1154290$ & $\text{Turn-off}$ & $6122$ & $3.82$ & $-0.07$ & $1.07$ & $10.32$ & $34.50$ \\
$6571679$ & $08512830+1152175$ & $\text{Turn-off}$ & $6121$ & $3.82$ & $-0.11$ & $1.07$ & $8.85$ & $34.49$ \\
$6567547$ & $08514082+1149055$ & $\text{Turn-off}$ & $6110$ & $4.30$ & $0.02$ & $1.05$ & $8.35$ & $34.54$ \\
$6555602$ & $08505973+1139524$ & $\text{Turn-off}$ & $6108$ & $4.05$ & $-0.10$ & $1.07$ & $7.07$ & $33.75$ \\
$6561039$ & $08514597+1144093$ & $\text{Turn-off}$ & $6106$ & $4.18$ & $-0.08$ & $1.06$ & $9.09$ & $33.75$ \\
$6570179$ & $08505474+1151093$ & $\text{Turn-off}$ & $6098$ & $4.16$ & $-0.15$ & $1.05$ & $7.40$ & $34.61$ \\
$6564123$ & $08514641+1146267$ & $\text{Turn-off}$ & $6094$ & $4.09$ & $-0.04$ & $1.05$ & $8.91$ & $34.53$ \\
$6573044$ & $08513119+1153179$ & $\text{Turn-off}$ & $6092$ & $4.28$ & $-0.07$ & $1.02$ & $7.37$ & $34.02$ \\
$6575508$ & $08505762+1155147$ & $\text{Turn-off}$ & $6088$ & $4.16$ & $-0.14$ & $1.03$ & $7.94$ & $34.19$ \\
$6558150$ & $08514465+1141510$ & $\text{Turn-off}$ & $6074$ & $3.98$ & $-0.12$ & $1.03$ & $9.18$ & $33.18$ \\
$6568768$ & $08513923+1150038$ & $\text{Turn-off}$ & $6071$ & $3.79$ & $-0.11$ & $1.04$ & $14.49$ & $36.38$ \\
$6573727$ & $08505600+1153520$ & $\text{Turn-off}$ & $6066$ & $4.16$ & $-0.11$ & $1.03$ & $7.85$ & $35.18$ \\
$6573191$ & $08512742+1153265$ & $\text{Turn-off}$ & $6062$ & $3.84$ & $-0.10$ & $1.04$ & $7.89$ & $33.98$ \\
$6572337$ & $08512015+1152479$ & $\text{Turn-off}$ & $6060$ & $4.24$ & $-0.03$ & $1.05$ & $7.83$ & $34.24$ \\
$6569861$ & $08510857+1150530$ & $\text{Turn-off}$ & $6058$ & $4.24$ & $-0.09$ & $1.03$ & $7.21$ & $35.74$ \\
$6564445$ & $08512205+1146409$ & $\text{Turn-off}$ & $6055$ & $3.93$ & $-0.08$ & $1.04$ & $9.63$ & $35.61$ \\
$6567617$ & $08512595+1149089$ & $\text{Turn-off}$ & $6052$ & $4.07$ & $-0.10$ & $1.02$ & $9.09$ & $35.12$ \\
$6559497$ & $08511810+1142547$ & $\text{Turn-off}$ & $6051$ & $4.05$ & $-0.03$ & $1.04$ & $7.12$ & $34.15$ \\
$6568479$ & $08520785+1149500$ & $\text{Turn-off}$ & $6050$ & $4.16$ & $-0.10$ & $1.04$ & $8.72$ & $33.72$ \\
$6572560$ & $08515963+1152576$ & $\text{Turn-off}$ & $6048$ & $3.69$ & $-0.07$ & $1.02$ & $8.48$ & $34.32$ \\
$6572187$ & $08512552+1152388$ & $\text{Turn-off}$ & $6046$ & $4.07$ & $-0.02$ & $1.03$ & $7.53$ & $34.95$ \\
$6560653$ & $08513012+1143498$ & $\text{Turn-off}$ & $6046$ & $4.24$ & $-0.09$ & $1.02$ & $8.19$ & $34.35$ \\
$6567233$ & $08511164+1148505$ & $\text{Turn-off}$ & $6040$ & $4.26$ & $-0.02$ & $1.02$ & $6.58$ & $36.38$ \\
$6569167$ & $08520741+1150221$ & $\text{Turn-off}$ & $6034$ & $3.94$ & $-0.05$ & $1.02$ & $8.23$ & $35.22$ \\
$6565967$ & $08510156+1147501$ & $\text{Turn-off}$ & $6032$ & $4.18$ & $-0.01$ & $1.02$ & $6.68$ & $33.59$ \\
$6562672$ & $08504760+1145228$ & $\text{Turn-off}$ & $6032$ & $4.24$ & $-0.16$ & $1.00$ & $6.46$ & $34.56$ \\
$6571851$ & $08510492+1152261$ & $\text{Turn-off}$ & $6025$ & $4.28$ & $-0.10$ & $1.01$ & $6.95$ & $35.13$ \\
$6568307$ & $08514914+1149435$ & $\text{Turn-off}$ & $6016$ & $3.76$ & $-0.06$ & $1.01$ & $8.38$ & $34.80$ \\
$8436138$ & $08504976+1154244$ & $\text{Turn-off}$ & $5995$ & $4.24$ & $-0.09$ & $0.99$ & $6.64$ & $33.53$ \\
$6571594$ & $08505569+1152146$ & $\text{Turn-off}$ & $5995$ & $3.77$ & $-0.04$ & $1.00$ & $7.53$ & $34.62$ \\
$6579199$ & $08520330+1158046$ & $\text{Turn-off}$ & $5980$ & $4.22$ & $0.01$ & $0.99$ & $6.64$ & $33.66$ \\
$6562188$ & $08512080+1145024$ & $\text{Turn-off}$ & $5942$ & $4.16$ & $-0.08$ & $0.96$ & $8.51$ & $35.09$ \\
$6567847$ & $08511854+1149214$ & $\text{Turn-off}$ & $5932$ & $3.74$ & $-0.06$ & $0.97$ & $7.35$ & $35.08$ \\
$6563234$ & $08510325+1145473$ & $\text{Turn-off}$ & $5929$ & $3.70$ & $-0.03$ & $0.98$ & $6.72$ & $35.63$ \\
$9077970$ & $08513540+1157564$ & $\text{Sub-giant}$ & $5563$ & $3.78$ & $-0.01$ & $0.89$ & $7.00$ & $33.71$ \\
$6568921$ & $08510106+1150108$ & $\text{Sub-giant}$ & $5558$ & $3.74$ & $-0.05$ & $0.89$ & $6.49$ & $33.23$ \\
$6574583$ & $08510018+1154321$ & $\text{Sub-giant}$ & $5441$ & $3.75$ & $0.04$ & $0.90$ & $7.18$ & $34.13$ \\
$6562991$ & $08521134+1145380$ & $\text{Sub-giant}$ & $5408$ & $3.75$ & $-0.03$ & $0.91$ & $7.40$ & $33.40$ \\
$6569862$ & $08511564+1150561$ & $\text{Sub-giant}$ & $5299$ & $3.76$ & $-0.04$ & $0.93$ & $6.16$ & $34.51$ \\
$6567693$ & $08504994+1149127$ & $\text{Sub-giant}$ & $5200$ & $3.61$ & $-0.15$ & $0.97$ & $6.19$ & $34.21$ \\
$6577630$ & $08514883+1156511$ & $\text{Sub-giant}$ & $5166$ & $3.68$ & $-0.05$ & $0.98$ & $8.01$ & $34.92$ \\
$6562765$ & $08512935+1145275$ & $\text{Sub-giant}$ & $5140$ & $3.55$ & $-0.07$ & $1.00$ & $6.74$ & $33.81$ \\
$6569012$ & $08515611+1150147$ & $\text{Sub-giant}$ & $5113$ & $3.69$ & $0.07$ & $1.04$ & $7.41$ & $35.89$ \\
$6571766$ & $08505816+1152223$ & $\text{Red-giant}$ & $5056$ & $3.55$ & $-0.08$ & $1.05$ & $5.85$ & $34.58$ \\
$6565104$ & $08510839+1147121$ & $\text{Red-giant}$ & $5029$ & $3.45$ & $-0.04$ & $1.06$ & $5.52$ & $34.27$ \\
$6579331$ & $08511897+1158110$ & $\text{Red-giant}$ & $5026$ & $3.44$ & $-0.04$ & $1.07$ & $6.47$ & $34.69$ \\
$6573364$ & $08513577+1153347$ & $\text{Red-giant}$ & $5020$ & $3.33$ & $-0.05$ & $1.07$ & $7.07$ & $34.76$ \\
$6563655$ & $08512156+1146061$ & $\text{Red-giant}$ & $4913$ & $3.13$ & $0.04$ & $1.17$ & $6.68$ & $35.28$ \\
$6570514$ & $08514235+1151230$ & $\text{Red-giant}$ & $4900$ & $3.25$ & $-0.02$ & $1.18$ & $7.66$ & $34.85$ \\
$6568851$ & $08514234+1150076$ & $\text{Red-giant}$ & $4898$ & $3.04$ & $-0.05$ & $1.17$ & $7.68$ & $34.73$ \\
$6565879$ & $08514507+1147459$ & $\text{Red-giant}$ & $4881$ & $3.06$ & $-0.07$ & $1.19$ & $7.49$ & $32.79$ \\
$6569711$ & $08511704+1150464$ & $\text{Red-giant}$ & $4839$ & $2.86$ & $-0.12$ & $1.24$ & $6.46$ & $34.35$ \\
$6575356$ & $08515952+1155049$ & $\text{Red-clump}$ & $4824$ & $2.62$ & $-0.08$ & $1.23$ & $7.78$ & $35.38$ \\
$6577481$ & $08514388+1156425$ & $\text{Red-clump}$ & $4822$ & $2.54$ & $-0.08$ & $1.21$ & $7.56$ & $33.82$ \\
$6573728$ & $08512618+1153520$ & $\text{Red-clump}$ & $4793$ & $2.55$ & $-0.10$ & $1.24$ & $6.98$ & $34.90$ \\
$6566179$ & $08512280+1148016$ & $\text{Red-clump}$ & $4787$ & $2.52$ & $-0.06$ & $1.25$ & $7.55$ & $34.23$ \\
$6569393$ & $08512898+1150330$ & $\text{Red-clump}$ & $4771$ & $2.67$ & $-0.08$ & $1.28$ & $7.48$ & $34.20$ \\
$6572270$ & $08511269+1152423$ & $\text{Red-clump}$ & $4761$ & $2.53$ & $-0.11$ & $1.24$ & $6.94$ & $35.14$ \\
\hline
\end{tabular}
\end{center}
\end{table*}

\section{Tables of the chemical abundances}
\begin{table*}
\begin{center}
\caption{Non-LTE chemical abundances of the sample stars in M67. Abundances were derived relative to non-LTE values of
solar analysed in this work.}
\label{tab:abund_err}
\begin{tabular}{c c c c c c c c c}
\hline
$\text{GALAH ID}$ &
$\text{Group}$ &
$\text{A(Li)}_{\mathrm{NLTE}}$ &
$\text{[O/H]}_{\mathrm{NLTE}}$ &
$\text{[Na/H]}_{\mathrm{NLTE}}$ &
$\text{[Mg/H]}_{\mathrm{NLTE}}$ &
$\text{[Al/H]}_{\mathrm{NLTE}}$ &
$\text{[Si/H]}_{\mathrm{NLTE}}$ &
$\text{[Fe/H]}_{\mathrm{NLTE}}$ \\
\hline
\hline
$6561552$ & $\text{Main-sequence}$ & $-$ & $0.15\pm0.12$ & $-0.08\pm0.04$ & $-0.02\pm0.04$ & $0.03\pm0.03$ & $-0.01\pm0.03$ & $-0.03\pm0.05$ \\
$6560101$ & $\text{Turn-off}$ & $2.68\pm0.05$ & $-0.02\pm0.12$ & $-0.07\pm0.04$ & $-0.06\pm0.03$ & $-0.22\pm0.03$ & $-0.03\pm0.03$ & $-0.04\pm0.04$ \\
$6577714$ & $\text{Turn-off}$ & $2.66\pm0.05$ & $-0.08\pm0.12$ & $-0.10\pm0.03$ & $-0.11\pm0.03$ & $-0.03\pm0.03$ & $-0.05\pm0.03$ & $-0.06\pm0.04$ \\
$6554484$ & $\text{Turn-off}$ & $2.68\pm0.05$ & $-0.09\pm0.12$ & $0.02\pm0.04$ & $-0.05\pm0.03$ & $-0.05\pm0.02$ & $-0.06\pm0.03$ & $0.04\pm0.04$ \\
$6577148$ & $\text{Turn-off}$ & $-$ & $-0.02\pm0.11$ & $-0.02\pm0.04$ & $-0.09\pm0.04$ & $-0.15\pm0.03$ & $-0.05\pm0.03$ & $-0.04\pm0.04$ \\
$6569011$ & $\text{Turn-off}$ & $2.34\pm0.05$ & $-0.03\pm0.12$ & $0.02\pm0.04$ & $-0.05\pm0.03$ & $-0.08\pm0.03$ & $-0.05\pm0.03$ & $-0.02\pm0.04$ \\
$6565966$ & $\text{Turn-off}$ & $2.54\pm0.05$ & $-0.08\pm0.12$ & $-0.05\pm0.04$ & $-0.09\pm0.03$ & $-0.04\pm0.03$ & $-0.07\pm0.03$ & $-0.04\pm0.04$ \\
$6565326$ & $\text{Turn-off}$ & $2.42\pm0.05$ & $-0.03\pm0.12$ & $0.04\pm0.04$ & $-0.06\pm0.03$ & $-0.03\pm0.02$ & $-0.03\pm0.03$ & $0.01\pm0.04$ \\
$6574584$ & $\text{Turn-off}$ & $-$ & $-0.09\pm0.12$ & $-0.02\pm0.04$ & $-0.02\pm0.03$ & $-0.11\pm0.03$ & $-0.07\pm0.03$ & $-0.02\pm0.04$ \\
$6571679$ & $\text{Turn-off}$ & $-$ & $-0.03\pm0.12$ & $0.00\pm0.04$ & $-0.06\pm0.04$ & $-0.17\pm0.03$ & $-0.07\pm0.03$ & $-0.06\pm0.04$ \\
$6567547$ & $\text{Turn-off}$ & $2.52\pm0.05$ & $-0.01\pm0.12$ & $-0.01\pm0.04$ & $-0.08\pm0.03$ & $-0.06\pm0.03$ & $-0.01\pm0.03$ & $0.03\pm0.04$ \\
$6555602$ & $\text{Turn-off}$ & $-$ & $-0.01\pm0.12$ & $-0.07\pm0.03$ & $-0.15\pm0.03$ & $-0.13\pm0.03$ & $-0.11\pm0.03$ & $-0.08\pm0.05$ \\
$6561039$ & $\text{Turn-off}$ & $2.70\pm0.05$ & $0.07\pm0.11$ & $-0.07\pm0.04$ & $-0.03\pm0.03$ & $-0.12\pm0.03$ & $-0.03\pm0.03$ & $-0.05\pm0.04$ \\
$6570179$ & $\text{Turn-off}$ & $2.44\pm0.05$ & $-0.02\pm0.11$ & $-0.13\pm0.03$ & $-0.15\pm0.03$ & $-0.24\pm0.03$ & $-0.14\pm0.03$ & $-0.14\pm0.05$ \\
$6564123$ & $\text{Turn-off}$ & $2.38\pm0.05$ & $-0.05\pm0.11$ & $-0.06\pm0.03$ & $-0.03\pm0.03$ & $-0.12\pm0.03$ & $-0.07\pm0.03$ & $-0.01\pm0.04$ \\
$6573044$ & $\text{Turn-off}$ & $2.45\pm0.05$ & $-0.03\pm0.12$ & $-0.05\pm0.04$ & $-0.10\pm0.03$ & $-0.14\pm0.03$ & $-0.05\pm0.03$ & $-0.03\pm0.05$ \\
$6575508$ & $\text{Turn-off}$ & $2.23\pm0.05$ & $-0.02\pm0.11$ & $-0.14\pm0.03$ & $-0.13\pm0.04$ & $-0.22\pm0.03$ & $-0.10\pm0.03$ & $-0.12\pm0.04$ \\
$6558150$ & $\text{Turn-off}$ & $2.52\pm0.05$ & $-0.03\pm0.12$ & $-0.02\pm0.04$ & $-0.12\pm0.03$ & $-0.15\pm0.02$ & $-0.08\pm0.03$ & $-0.07\pm0.04$ \\
$6568768$ & $\text{Turn-off}$ & $-$ & $0.00\pm0.12$ & $-0.03\pm0.04$ & $-0.03\pm0.03$ & $-0.09\pm0.03$ & $-0.09\pm0.03$ & $-0.05\pm0.05$ \\
$6573727$ & $\text{Turn-off}$ & $2.47\pm0.05$ & $-0.09\pm0.12$ & $-0.11\pm0.03$ & $-0.14\pm0.03$ & $-0.09\pm0.03$ & $-0.07\pm0.03$ & $-0.08\pm0.04$ \\
$6573191$ & $\text{Turn-off}$ & $-$ & $-0.02\pm0.11$ & $-0.01\pm0.04$ & $-0.11\pm0.03$ & $-0.11\pm0.03$ & $-0.10\pm0.02$ & $-0.06\pm0.04$ \\
$6572337$ & $\text{Turn-off}$ & $2.35\pm0.05$ & $-0.05\pm0.12$ & $-0.02\pm0.04$ & $-0.08\pm0.04$ & $0.05\pm0.03$ & $-0.05\pm0.03$ & $-0.03\pm0.04$ \\
$6569861$ & $\text{Turn-off}$ & $2.26\pm0.05$ & $-0.03\pm0.11$ & $-0.10\pm0.04$ & $-0.12\pm0.03$ & $-0.22\pm0.03$ & $-0.10\pm0.03$ & $-0.06\pm0.04$ \\
$6564445$ & $\text{Turn-off}$ & $2.22\pm0.05$ & $-0.16\pm0.12$ & $0.00\pm0.04$ & $0.00\pm0.04$ & $-0.17\pm0.03$ & $-0.03\pm0.02$ & $-0.06\pm0.04$ \\
$6567617$ & $\text{Turn-off}$ & $2.68\pm0.05$ & $-0.03\pm0.12$ & $-0.05\pm0.04$ & $-0.07\pm0.04$ & $-0.05\pm0.03$ & $-0.04\pm0.03$ & $-0.07\pm0.04$ \\
$6559497$ & $\text{Turn-off}$ & $2.58\pm0.05$ & $-0.06\pm0.12$ & $0.02\pm0.04$ & $-0.05\pm0.04$ & $-0.10\pm0.02$ & $-0.01\pm0.02$ & $0.00\pm0.04$ \\
$6568479$ & $\text{Turn-off}$ & $1.71\pm0.05$ & $-0.09\pm0.12$ & $-0.06\pm0.04$ & $-0.09\pm0.03$ & $-0.11\pm0.02$ & $-0.08\pm0.03$ & $-0.08\pm0.04$ \\
$6572560$ & $\text{Turn-off}$ & $2.51\pm0.05$ & $-0.06\pm0.12$ & $0.04\pm0.03$ & $-0.02\pm0.03$ & $-0.09\pm0.03$ & $-0.06\pm0.03$ & $-0.02\pm0.04$ \\
$6572187$ & $\text{Turn-off}$ & $2.06\pm0.05$ & $0.02\pm0.12$ & $0.00\pm0.04$ & $-0.05\pm0.03$ & $-0.07\pm0.03$ & $-0.04\pm0.03$ & $0.00\pm0.04$ \\
$6560653$ & $\text{Turn-off}$ & $2.43\pm0.05$ & $-0.11\pm0.12$ & $-0.08\pm0.04$ & $-0.12\pm0.03$ & $-0.06\pm0.03$ & $-0.10\pm0.03$ & $-0.06\pm0.04$ \\
$6567233$ & $\text{Turn-off}$ & $2.65\pm0.05$ & $0.02\pm0.12$ & $0.02\pm0.04$ & $-0.02\pm0.03$ & $-0.07\pm0.02$ & $-0.03\pm0.03$ & $-0.01\pm0.05$ \\
$6569167$ & $\text{Turn-off}$ & $2.45\pm0.05$ & $-0.05\pm0.12$ & $-0.06\pm0.03$ & $-0.04\pm0.03$ & $-0.15\pm0.03$ & $-0.04\pm0.03$ & $-0.02\pm0.04$ \\
$6565967$ & $\text{Turn-off}$ & $2.62\pm0.05$ & $-0.01\pm0.12$ & $0.02\pm0.04$ & $-0.06\pm0.03$ & $-0.03\pm0.03$ & $-0.03\pm0.03$ & $0.02\pm0.04$ \\
$6562672$ & $\text{Turn-off}$ & $2.42\pm0.05$ & $-0.05\pm0.12$ & $-0.17\pm0.03$ & $-0.15\pm0.03$ & $-0.14\pm0.03$ & $-0.17\pm0.03$ & $-0.13\pm0.04$ \\
$6571851$ & $\text{Turn-off}$ & $2.34\pm0.05$ & $-0.06\pm0.11$ & $-0.09\pm0.04$ & $-0.12\pm0.03$ & $-0.12\pm0.03$ & $-0.06\pm0.03$ & $-0.08\pm0.04$ \\
$6568307$ & $\text{Turn-off}$ & $-$ & $-0.05\pm0.12$ & $0.01\pm0.04$ & $0.00\pm0.04$ & $-0.08\pm0.03$ & $-0.04\pm0.03$ & $-0.02\pm0.04$ \\
$8436138$ & $\text{Turn-off}$ & $2.34\pm0.05$ & $-0.06\pm0.12$ & $-0.08\pm0.03$ & $-0.11\pm0.03$ & $-0.12\pm0.04$ & $-0.08\pm0.03$ & $-0.05\pm0.04$ \\
$6571594$ & $\text{Turn-off}$ & $2.13\pm0.05$ & $-0.01\pm0.12$ & $-0.02\pm0.03$ & $-0.01\pm0.03$ & $0.04\pm0.03$ & $-0.03\pm0.02$ & $-0.01\pm0.04$ \\
$6579199$ & $\text{Turn-off}$ & $2.21\pm0.05$ & $-0.02\pm0.12$ & $-0.10\pm0.03$ & $-0.06\pm0.03$ & $-0.02\pm0.03$ & $-0.06\pm0.03$ & $0.03\pm0.05$ \\
$6562188$ & $\text{Turn-off}$ & $-$ & $0.12\pm0.11$ & $-0.07\pm0.04$ & $-0.05\pm0.03$ & $-0.16\pm0.02$ & $-0.11\pm0.03$ & $-0.05\pm0.05$ \\
$6567847$ & $\text{Turn-off}$ & $-$ & $-0.01\pm0.12$ & $0.01\pm0.03$ & $-0.08\pm0.03$ & $-0.10\pm0.05$ & $-0.04\pm0.02$ & $-0.02\pm0.04$ \\
$6563234$ & $\text{Turn-off}$ & $-$ & $-0.01\pm0.12$ & $0.05\pm0.04$ & $0.05\pm0.04$ & $-0.06\pm0.04$ & $-0.03\pm0.03$ & $0.00\pm0.04$ \\
$9077970$ & $\text{Sub-giant}$ & $-$ & $0.12\pm0.13$ & $0.02\pm0.04$ & $0.00\pm0.03$ & $-0.07\pm0.02$ & $0.03\pm0.03$ & $0.02\pm0.04$ \\
$6568921$ & $\text{Sub-giant}$ & $-$ & $0.05\pm0.12$ & $0.01\pm0.04$ & $0.02\pm0.03$ & $-0.01\pm0.02$ & $0.00\pm0.03$ & $-0.02\pm0.04$ \\
$6574583$ & $\text{Sub-giant}$ & $-$ & $-0.01\pm0.13$ & $0.08\pm0.04$ & $0.08\pm0.03$ & $0.10\pm0.03$ & $0.04\pm0.03$ & $0.07\pm0.04$ \\
$6562991$ & $\text{Sub-giant}$ & $-$ & $0.00\pm0.13$ & $-0.04\pm0.04$ & $0.01\pm0.04$ & $-0.04\pm0.03$ & $0.00\pm0.03$ & $-0.01\pm0.04$ \\
$6569862$ & $\text{Sub-giant}$ & $-$ & $0.01\pm0.13$ & $-0.04\pm0.04$ & $-0.05\pm0.04$ & $0.01\pm0.03$ & $0.03\pm0.04$ & $-0.02\pm0.04$ \\
$6567693$ & $\text{Sub-giant}$ & $-$ & $0.12\pm0.13$ & $-0.14\pm0.04$ & $-0.08\pm0.04$ & $-0.09\pm0.03$ & $-0.10\pm0.04$ & $-0.12\pm0.04$ \\
$6577630$ & $\text{Sub-giant}$ & $-$ & $-0.14\pm0.13$ & $-0.02\pm0.05$ & $-0.08\pm0.04$ & $-0.02\pm0.02$ & $-0.07\pm0.04$ & $-0.05\pm0.04$ \\
$6562765$ & $\text{Sub-giant}$ & $-$ & $-0.15\pm0.13$ & $-0.06\pm0.05$ & $-0.02\pm0.04$ & $-0.02\pm0.03$ & $-0.04\pm0.04$ & $-0.05\pm0.04$ \\
$6569012$ & $\text{Sub-giant}$ & $-$ & $-0.15\pm0.14$ & $0.06\pm0.05$ & $0.17\pm0.04$ & $0.10\pm0.03$ & $0.02\pm0.05$ & $0.07\pm0.04$ \\
$6571766$ & $\text{Red-giant}$ & $-$ & $0.04\pm0.14$ & $-0.08\pm0.04$ & $0.01\pm0.04$ & $0.02\pm0.02$ & $-0.06\pm0.05$ & $-0.07\pm0.05$ \\
$6565104$ & $\text{Red-giant}$ & $-$ & $-0.09\pm0.14$ & $-0.04\pm0.04$ & $0.00\pm0.04$ & $0.06\pm0.03$ & $0.01\pm0.05$ & $-0.03\pm0.05$ \\
$6579331$ & $\text{Red-giant}$ & $-$ & $0.20\pm0.14$ & $-0.05\pm0.06$ & $-0.02\pm0.04$ & $0.02\pm0.03$ & $0.00\pm0.04$ & $-0.04\pm0.04$ \\
$6573364$ & $\text{Red-giant}$ & $-$ & $-0.05\pm0.14$ & $-0.03\pm0.05$ & $-0.03\pm0.04$ & $-0.01\pm0.03$ & $-0.01\pm0.04$ & $-0.01\pm0.05$ \\
$6563655$ & $\text{Red-giant}$ & $-$ & $-0.04\pm0.15$ & $0.02\pm0.04$ & $0.11\pm0.04$ & $0.18\pm0.03$ & $0.05\pm0.05$ & $0.03\pm0.05$ \\
$6570514$ & $\text{Red-giant}$ & $-$ & $-0.08\pm0.14$ & $-0.05\pm0.04$ & $-0.11\pm0.04$ & $-0.02\pm0.03$ & $0.01\pm0.05$ & $-0.06\pm0.04$ \\
$6568851$ & $\text{Red-giant}$ & $-$ & $0.00\pm0.15$ & $-0.09\pm0.05$ & $-0.12\pm0.04$ & $0.06\pm0.03$ & $-0.05\pm0.05$ & $-0.07\pm0.05$ \\
$6565879$ & $\text{Red-giant}$ & $-$ & $0.07\pm0.15$ & $-0.09\pm0.05$ & $-0.15\pm0.04$ & $0.10\pm0.03$ & $-0.01\pm0.05$ & $-0.09\pm0.04$ \\
$6569711$ & $\text{Red-giant}$ & $-$ & $0.00\pm0.14$ & $-0.05\pm0.04$ & $-0.19\pm0.04$ & $0.00\pm0.04$ & $-0.10\pm0.05$ & $-0.13\pm0.05$ \\
$6575356$ & $\text{Red-clump}$ & $-$ & $-0.06\pm0.14$ & $0.04\pm0.05$ & $-0.12\pm0.04$ & $0.05\pm0.03$ & $-0.03\pm0.05$ & $-0.08\pm0.05$ \\
$6577481$ & $\text{Red-clump}$ & $-$ & $-0.22\pm0.15$ & $0.05\pm0.04$ & $-0.09\pm0.05$ & $0.01\pm0.03$ & $-0.03\pm0.05$ & $-0.07\pm0.05$ \\
$6573728$ & $\text{Red-clump}$ & $-$ & $-0.13\pm0.15$ & $0.05\pm0.05$ & $-0.08\pm0.04$ & $0.00\pm0.04$ & $-0.03\pm0.04$ & $-0.09\pm0.05$ \\
$6566179$ & $\text{Red-clump}$ & $-$ & $0.12\pm0.15$ & $0.12\pm0.05$ & $-0.03\pm0.05$ & $0.06\pm0.03$ & $0.02\pm0.05$ & $-0.05\pm0.05$ \\
$6569393$ & $\text{Red-clump}$ & $-$ & $0.07\pm0.15$ & $-0.07\pm0.04$ & $-0.14\pm0.05$ & $0.00\pm0.03$ & $-0.01\pm0.05$ & $-0.10\pm0.05$ \\
$6572270$ & $\text{Red-clump}$ & $-$ & $0.00\pm0.15$ & $0.06\pm0.05$ & $-0.13\pm0.04$ & $0.03\pm0.03$ & $-0.01\pm0.05$ & $-0.08\pm0.05$ \\
\hline
\end{tabular}
\end{center}
\end{table*}

\begin{table*}
\begin{center}
\caption{LTE chemical abundances of the sample stars in M67. Abundances were derived relative to non-LTE values of
solar analysed in this work.}
\label{tab:abund_err}
\begin{tabular}{c c c c c c c c c}
\hline
$\text{GALAH ID}$ &
$\text{Group}$ &
$\text{A(Li)}_{\mathrm{LTE}}$ &
$\text{[O/H]}_{\mathrm{LTE}}$ &
$\text{[Na/H]}_{\mathrm{LTE}}$ &
$\text{[Mg/H]}_{\mathrm{LTE}}$ &
$\text{[Al/H]}_{\mathrm{LTE}}$ &
$\text{[Si/H]}_{\mathrm{LTE}}$ &
$\text{[Fe/H]}_{\mathrm{LTE}}$ \\
\hline
\hline
$6561552$ & $\text{Main-sequence}$ & $-$ & $0.33\pm0.10$ & $0.07\pm0.05$ & $-0.03\pm0.04$ & $0.06\pm0.02$ & $0.00\pm0.03$ & $-0.02\pm0.05$ \\
$6560101$ & $\text{Turn-off}$ & $2.68\pm0.05$ & $0.32\pm0.10$ & $0.11\pm0.04$ & $-0.09\pm0.03$ & $-0.21\pm0.02$ & $-0.02\pm0.03$ & $-0.03\pm0.04$ \\
$6577714$ & $\text{Turn-off}$ & $2.66\pm0.05$ & $0.24\pm0.10$ & $0.07\pm0.04$ & $-0.14\pm0.03$ & $-0.02\pm0.02$ & $-0.04\pm0.03$ & $-0.05\pm0.05$ \\
$6554484$ & $\text{Turn-off}$ & $2.68\pm0.05$ & $0.22\pm0.10$ & $0.19\pm0.04$ & $-0.07\pm0.03$ & $-0.04\pm0.02$ & $-0.05\pm0.03$ & $0.05\pm0.04$ \\
$6577148$ & $\text{Turn-off}$ & $-$ & $0.37\pm0.10$ & $0.16\pm0.04$ & $-0.13\pm0.03$ & $-0.14\pm0.02$ & $-0.04\pm0.03$ & $-0.03\pm0.04$ \\
$6569011$ & $\text{Turn-off}$ & $2.34\pm0.05$ & $0.37\pm0.10$ & $0.20\pm0.04$ & $-0.10\pm0.03$ & $-0.07\pm0.02$ & $-0.04\pm0.03$ & $0.00\pm0.04$ \\
$6565966$ & $\text{Turn-off}$ & $2.54\pm0.05$ & $0.22\pm0.10$ & $0.13\pm0.04$ & $-0.11\pm0.03$ & $-0.03\pm0.03$ & $-0.07\pm0.03$ & $-0.03\pm0.04$ \\
$6565326$ & $\text{Turn-off}$ & $2.42\pm0.05$ & $0.37\pm0.10$ & $0.23\pm0.04$ & $-0.10\pm0.03$ & $-0.02\pm0.02$ & $-0.02\pm0.03$ & $0.03\pm0.04$ \\
$6574584$ & $\text{Turn-off}$ & $-$ & $0.29\pm0.11$ & $0.16\pm0.04$ & $-0.06\pm0.03$ & $-0.11\pm0.02$ & $-0.06\pm0.03$ & $0.00\pm0.04$ \\
$6571679$ & $\text{Turn-off}$ & $-$ & $0.38\pm0.10$ & $0.19\pm0.03$ & $-0.10\pm0.03$ & $-0.17\pm0.02$ & $-0.06\pm0.03$ & $-0.04\pm0.04$ \\
$6567547$ & $\text{Turn-off}$ & $2.52\pm0.05$ & $0.26\pm0.10$ & $0.16\pm0.04$ & $-0.10\pm0.03$ & $-0.05\pm0.02$ & $0.00\pm0.02$ & $0.04\pm0.04$ \\
$6555602$ & $\text{Turn-off}$ & $-$ & $0.32\pm0.10$ & $0.11\pm0.04$ & $-0.18\pm0.03$ & $-0.13\pm0.02$ & $-0.10\pm0.03$ & $-0.07\pm0.04$ \\
$6561039$ & $\text{Turn-off}$ & $2.70\pm0.05$ & $0.39\pm0.10$ & $0.07\pm0.04$ & $-0.06\pm0.03$ & $-0.11\pm0.02$ & $-0.02\pm0.03$ & $-0.05\pm0.05$ \\
$6570179$ & $\text{Turn-off}$ & $2.44\pm0.05$ & $0.28\pm0.10$ & $0.04\pm0.04$ & $-0.18\pm0.03$ & $-0.23\pm0.03$ & $-0.13\pm0.03$ & $-0.14\pm0.04$ \\
$6564123$ & $\text{Turn-off}$ & $2.38\pm0.05$ & $0.26\pm0.10$ & $0.11\pm0.04$ & $-0.05\pm0.03$ & $-0.11\pm0.02$ & $-0.06\pm0.02$ & $0.00\pm0.05$ \\
$6573044$ & $\text{Turn-off}$ & $2.45\pm0.05$ & $0.24\pm0.10$ & $0.13\pm0.04$ & $-0.12\pm0.03$ & $-0.13\pm0.03$ & $-0.04\pm0.03$ & $-0.03\pm0.04$ \\
$6575508$ & $\text{Turn-off}$ & $2.23\pm0.05$ & $0.29\pm0.10$ & $0.03\pm0.04$ & $-0.16\pm0.03$ & $-0.21\pm0.02$ & $-0.08\pm0.03$ & $-0.11\pm0.05$ \\
$6558150$ & $\text{Turn-off}$ & $2.52\pm0.05$ & $0.30\pm0.10$ & $0.17\pm0.04$ & $-0.15\pm0.03$ & $-0.14\pm0.02$ & $-0.07\pm0.03$ & $-0.06\pm0.05$ \\
$6568768$ & $\text{Turn-off}$ & $-$ & $0.39\pm0.10$ & $0.14\pm0.04$ & $-0.07\pm0.03$ & $-0.09\pm0.03$ & $-0.08\pm0.03$ & $-0.04\pm0.04$ \\
$6573727$ & $\text{Turn-off}$ & $2.47\pm0.05$ & $0.18\pm0.10$ & $0.05\pm0.04$ & $-0.17\pm0.03$ & $-0.08\pm0.02$ & $-0.06\pm0.03$ & $-0.07\pm0.04$ \\
$6573191$ & $\text{Turn-off}$ & $-$ & $0.35\pm0.10$ & $0.18\pm0.04$ & $-0.14\pm0.04$ & $-0.10\pm0.02$ & $-0.09\pm0.02$ & $-0.05\pm0.05$ \\
$6572337$ & $\text{Turn-off}$ & $2.35\pm0.05$ & $0.21\pm0.11$ & $0.16\pm0.04$ & $-0.10\pm0.03$ & $0.06\pm0.03$ & $-0.04\pm0.03$ & $-0.01\pm0.04$ \\
$6569861$ & $\text{Turn-off}$ & $2.26\pm0.05$ & $0.24\pm0.10$ & $0.06\pm0.04$ & $-0.15\pm0.03$ & $-0.21\pm0.02$ & $-0.09\pm0.02$ & $-0.05\pm0.04$ \\
$6564445$ & $\text{Turn-off}$ & $2.22\pm0.05$ & $0.13\pm0.10$ & $0.19\pm0.04$ & $-0.03\pm0.03$ & $-0.16\pm0.03$ & $-0.02\pm0.02$ & $-0.05\pm0.04$ \\
$6567617$ & $\text{Turn-off}$ & $2.68\pm0.05$ & $0.27\pm0.11$ & $0.14\pm0.04$ & $-0.10\pm0.03$ & $-0.04\pm0.03$ & $-0.02\pm0.03$ & $-0.06\pm0.05$ \\
$6559497$ & $\text{Turn-off}$ & $2.58\pm0.05$ & $0.24\pm0.10$ & $0.20\pm0.04$ & $-0.08\pm0.03$ & $-0.08\pm0.02$ & $0.00\pm0.03$ & $0.01\pm0.05$ \\
$6568479$ & $\text{Turn-off}$ & $1.71\pm0.05$ & $0.17\pm0.11$ & $0.07\pm0.04$ & $-0.12\pm0.03$ & $-0.10\pm0.03$ & $-0.07\pm0.03$ & $-0.07\pm0.04$ \\
$6572560$ & $\text{Turn-off}$ & $2.50\pm0.05$ & $0.36\pm0.11$ & $0.23\pm0.04$ & $-0.07\pm0.03$ & $-0.08\pm0.03$ & $-0.04\pm0.03$ & $0.00\pm0.04$ \\
$6572187$ & $\text{Turn-off}$ & $2.05\pm0.05$ & $0.30\pm0.10$ & $0.18\pm0.04$ & $-0.07\pm0.03$ & $-0.05\pm0.03$ & $-0.02\pm0.03$ & $0.01\pm0.05$ \\
$6560653$ & $\text{Turn-off}$ & $2.43\pm0.05$ & $0.13\pm0.10$ & $0.09\pm0.04$ & $-0.14\pm0.03$ & $-0.04\pm0.02$ & $-0.09\pm0.03$ & $-0.05\pm0.05$ \\
$6567233$ & $\text{Turn-off}$ & $2.65\pm0.05$ & $0.28\pm0.10$ & $0.19\pm0.05$ & $-0.04\pm0.03$ & $-0.05\pm0.02$ & $-0.02\pm0.03$ & $0.00\pm0.04$ \\
$6569167$ & $\text{Turn-off}$ & $2.45\pm0.05$ & $0.28\pm0.11$ & $0.12\pm0.04$ & $-0.08\pm0.03$ & $-0.14\pm0.03$ & $-0.02\pm0.03$ & $0.00\pm0.04$ \\
$6565967$ & $\text{Turn-off}$ & $2.62\pm0.05$ & $0.28\pm0.11$ & $0.20\pm0.04$ & $-0.08\pm0.03$ & $-0.01\pm0.02$ & $-0.02\pm0.02$ & $0.03\pm0.04$ \\
$6562672$ & $\text{Turn-off}$ & $2.42\pm0.05$ & $0.20\pm0.10$ & $-0.01\pm0.04$ & $-0.17\pm0.03$ & $-0.13\pm0.03$ & $-0.16\pm0.02$ & $-0.12\pm0.05$ \\
$6571851$ & $\text{Turn-off}$ & $2.34\pm0.05$ & $0.17\pm0.10$ & $0.08\pm0.04$ & $-0.14\pm0.03$ & $-0.11\pm0.02$ & $-0.04\pm0.03$ & $-0.06\pm0.05$ \\
$6568307$ & $\text{Turn-off}$ & $-$ & $0.32\pm0.11$ & $0.21\pm0.04$ & $-0.06\pm0.03$ & $-0.07\pm0.02$ & $-0.03\pm0.03$ & $0.01\pm0.04$ \\
$8436138$ & $\text{Turn-off}$ & $2.34\pm0.05$ & $0.18\pm0.10$ & $0.10\pm0.04$ & $-0.12\pm0.04$ & $-0.10\pm0.03$ & $-0.06\pm0.02$ & $-0.04\pm0.05$ \\
$6571594$ & $\text{Turn-off}$ & $2.12\pm0.05$ & $0.37\pm0.10$ & $0.18\pm0.04$ & $-0.05\pm0.03$ & $0.06\pm0.02$ & $-0.01\pm0.02$ & $0.02\pm0.05$ \\
$6579199$ & $\text{Turn-off}$ & $2.21\pm0.05$ & $0.23\pm0.10$ & $0.06\pm0.04$ & $-0.09\pm0.03$ & $0.00\pm0.03$ & $-0.05\pm0.03$ & $0.04\pm0.04$ \\
$6562188$ & $\text{Turn-off}$ & $-$ & $0.40\pm0.10$ & $0.10\pm0.04$ & $-0.07\pm0.03$ & $-0.14\pm0.02$ & $-0.09\pm0.02$ & $-0.04\pm0.05$ \\
$6567847$ & $\text{Turn-off}$ & $-$ & $0.35\pm0.11$ & $0.20\pm0.04$ & $-0.12\pm0.04$ & $-0.07\pm0.04$ & $-0.03\pm0.03$ & $0.00\pm0.04$ \\
$6563234$ & $\text{Turn-off}$ & $-$ & $0.35\pm0.11$ & $0.26\pm0.04$ & $0.00\pm0.03$ & $-0.04\pm0.03$ & $-0.01\pm0.03$ & $0.03\pm0.04$ \\
$9077970$ & $\text{Sub-giant}$ & $-$ & $0.36\pm0.12$ & $0.21\pm0.04$ & $-0.02\pm0.03$ & $-0.02\pm0.02$ & $0.05\pm0.03$ & $0.05\pm0.04$ \\
$6568921$ & $\text{Sub-giant}$ & $-$ & $0.29\pm0.12$ & $0.20\pm0.04$ & $0.00\pm0.03$ & $0.03\pm0.02$ & $0.01\pm0.03$ & $0.01\pm0.05$ \\
$6574583$ & $\text{Sub-giant}$ & $-$ & $0.17\pm0.12$ & $0.25\pm0.05$ & $0.07\pm0.03$ & $0.15\pm0.03$ & $0.06\pm0.03$ & $0.11\pm0.05$ \\
$6562991$ & $\text{Sub-giant}$ & $-$ & $0.17\pm0.12$ & $0.13\pm0.05$ & $0.00\pm0.04$ & $0.01\pm0.03$ & $0.02\pm0.03$ & $0.02\pm0.05$ \\
$6569862$ & $\text{Sub-giant}$ & $-$ & $0.16\pm0.13$ & $0.13\pm0.05$ & $-0.05\pm0.03$ & $0.06\pm0.02$ & $0.05\pm0.03$ & $0.01\pm0.05$ \\
$6567693$ & $\text{Sub-giant}$ & $-$ & $0.29\pm0.12$ & $0.03\pm0.05$ & $-0.07\pm0.03$ & $-0.03\pm0.03$ & $-0.09\pm0.04$ & $-0.09\pm0.05$ \\
$6577630$ & $\text{Sub-giant}$ & $-$ & $-0.02\pm0.13$ & $0.14\pm0.05$ & $-0.07\pm0.03$ & $0.04\pm0.03$ & $-0.06\pm0.04$ & $-0.02\pm0.05$ \\
$6562765$ & $\text{Sub-giant}$ & $-$ & $-0.03\pm0.13$ & $0.10\pm0.05$ & $-0.01\pm0.03$ & $0.05\pm0.03$ & $-0.02\pm0.04$ & $-0.02\pm0.04$ \\
$6569012$ & $\text{Sub-giant}$ & $-$ & $-0.03\pm0.13$ & $0.24\pm0.05$ & $0.18\pm0.03$ & $0.16\pm0.03$ & $0.04\pm0.04$ & $0.10\pm0.05$ \\
$6571766$ & $\text{Red-giant}$ & $-$ & $0.17\pm0.13$ & $0.08\pm0.07$ & $0.03\pm0.04$ & $0.09\pm0.03$ & $-0.04\pm0.04$ & $-0.03\pm0.05$ \\
$6565104$ & $\text{Red-giant}$ & $-$ & $0.04\pm0.13$ & $0.14\pm0.06$ & $0.02\pm0.04$ & $0.14\pm0.03$ & $0.03\pm0.05$ & $0.04\pm0.05$ \\
$6579331$ & $\text{Red-giant}$ & $-$ & $0.37\pm0.13$ & $0.14\pm0.06$ & $0.00\pm0.04$ & $0.10\pm0.03$ & $0.02\pm0.05$ & $0.01\pm0.05$ \\
$6573364$ & $\text{Red-giant}$ & $-$ & $0.09\pm0.13$ & $0.18\pm0.06$ & $-0.02\pm0.04$ & $0.07\pm0.03$ & $0.01\pm0.04$ & $0.05\pm0.06$ \\
$6563655$ & $\text{Red-giant}$ & $-$ & $0.10\pm0.14$ & $0.23\pm0.05$ & $0.12\pm0.04$ & $0.27\pm0.03$ & $0.07\pm0.04$ & $0.09\pm0.05$ \\
$6570514$ & $\text{Red-giant}$ & $-$ & $0.05\pm0.14$ & $0.15\pm0.05$ & $-0.09\pm0.04$ & $0.05\pm0.04$ & $0.04\pm0.04$ & $-0.01\pm0.05$ \\
$6568851$ & $\text{Red-giant}$ & $-$ & $0.15\pm0.14$ & $0.13\pm0.06$ & $-0.11\pm0.04$ & $0.15\pm0.03$ & $-0.02\pm0.05$ & $-0.02\pm0.05$ \\
$6565879$ & $\text{Red-giant}$ & $-$ & $0.21\pm0.14$ & $0.15\pm0.06$ & $-0.13\pm0.04$ & $0.19\pm0.03$ & $0.01\pm0.05$ & $-0.02\pm0.04$ \\
$6569711$ & $\text{Red-giant}$ & $-$ & $0.14\pm0.14$ & $0.18\pm0.06$ & $-0.17\pm0.04$ & $0.10\pm0.04$ & $-0.07\pm0.05$ & $-0.06\pm0.05$ \\
$6575356$ & $\text{Red-clump}$ & $-$ & $0.13\pm0.15$ & $0.28\pm0.05$ & $-0.12\pm0.04$ & $0.15\pm0.03$ & $0.01\pm0.05$ & $0.00\pm0.05$ \\
$6577481$ & $\text{Red-clump}$ & $-$ & $-0.07\pm0.15$ & $0.29\pm0.05$ & $-0.10\pm0.04$ & $0.12\pm0.03$ & $0.01\pm0.04$ & $0.01\pm0.05$ \\
$6573728$ & $\text{Red-clump}$ & $-$ & $0.05\pm0.14$ & $0.28\pm0.05$ & $-0.07\pm0.03$ & $0.10\pm0.03$ & $0.00\pm0.05$ & $-0.01\pm0.05$ \\
$6566179$ & $\text{Red-clump}$ & $-$ & $0.29\pm0.14$ & $0.34\pm0.05$ & $-0.03\pm0.04$ & $0.18\pm0.03$ & $0.05\pm0.05$ & $0.04\pm0.05$ \\
$6569393$ & $\text{Red-clump}$ & $-$ & $0.25\pm0.15$ & $0.16\pm0.05$ & $-0.13\pm0.04$ & $0.10\pm0.03$ & $0.03\pm0.05$ & $-0.04\pm0.05$ \\
$6572270$ & $\text{Red-clump}$ & $-$ & $0.16\pm0.15$ & $0.28\pm0.05$ & $-0.12\pm0.04$ & $0.15\pm0.04$ & $0.03\pm0.05$ & $0.00\pm0.05$ \\
\hline
\end{tabular}
\end{center}
\end{table*}

\end{appendix}
\label{lastpage}
\end{document}